\documentclass[a4paper]{jpconf}
\usepackage{graphicx,amsmath,amssymb}

\usepackage{cite}
\newcommand{\abs}[1]{\lvert#1\rvert}
\newcommand{\ro}{\pmb{\rho}}
\newcommand{\lbar}{\overline{l}}

\newcommand{\rr}{\mathbf{r}}

\newcommand{\ess}{\mathbf{s}}

\newcommand{\kk}{\mathbf{k}}

\newcommand{\qq}{\mathbf{q}}

\newcommand{\ii}{\mathrm{i}}

\let \Im \relax
\DeclareMathOperator{\Im}{Im}

\begin{document}

\title{A Green's function decoupling scheme for the Edwards fermion-boson model}

\author{ D M Edwards$^1$,  S Ejima$^2$, A Alvermann$^{2}$, and H Fehske$^2$}

\address{$^1$Department of Mathematics, Imperial College London, London SW7~2BZ, United Kingdom}

\address{$^2$Institute of Physics, Ernst Moritz Arndt University, D-17489 
Greifswald, Germany}


\ead{d.edwards@imperial.ac.uk}

\begin{abstract}Holes in a Mott insulator are represented by spinless fermions in the fermion-boson model introduced by Edwards. Although the physically interesting regime is for low to moderate fermion density the model has interesting properties over the whole density range. It has previously been studied at half-filling in the one-dimensional (1D) case by numerical methods, in particular exact diagonalization and density matrix renormalization group (DMRG). In the present study the one-particle Green's function is calculated analytically by means of a decoupling scheme for the equations of motion, valid for arbitrary density in 1D, 2D and 3D with fairly large boson energy and zero boson relaxation parameter. The Green's function is used to compute some ground state properties, and the one-fermion spectral function, for fermion densities n=0.1, 0.5 and 0.9 in the 1D case. The results are generally in good agreement with numerical results obtained by DMRG and dynamical DMRG and new light is shed on the nature of the 
ground state at different fillings. The Green's function approximation is sufficiently successful in 1D to justify future application to the 2D and 3D cases. 
\end{abstract}

\section{Introduction}\label{sec:intro}
Hubbard-like models provide a paradigm for a large class of strongly correlated systems. A general form for the Hubbard Hamiltonian is
\begin{equation}\label{HamHub}
{\cal H}_{\rm Hu} ={\cal T}+{\cal U}=- \sum_{\mathbf{r}
,\ro,a,b} t_{\ro ab} c_{\mathbf{r}+\ro b}^{\dagger} c_{\mathbf{r}a}^{}   \\
+ U \sum_\mathbf{r} n_{\mathbf{r}+} n_{\mathbf{r}-} 
\end{equation}
and the system is strongly correlated when the repulsive on-site interaction $U$ is considerably larger than the hopping parameters $t$. Here $c_{\mathbf{r}a}$ destroys an electron in state $a$ on lattice site $\mathbf{r}$ and
 $c_{\mathbf{r}+\ro b}^{\dagger}$
creates an electron in state $b$ on a nearest-neighbour site $\mathbf{r}+\ro$. The state indices $a,b$ are summed over two states denoted by $+$ and $-$, and the occupation numbers $n_{\mathbf{r}\pm}= c_{\mathbf{r}\pm}^{\dagger} c_{\mathbf{r}\pm}$. In general we consider bipartite lattices in one, two and three dimensions. The standard Hubbard model~\cite{Hu63} corresponds to the case of a single orbital on each site with states $\pm$ corresponding to spin $\pm1/2$ and with hopping parameter
 \begin{equation}\label{HubHop}
  t_{\ro ab}=t_{0}\delta_{ab}\,.
   \end{equation}
   This model with a 2D square lattice is frequently used to describe the copper-oxygen plane of high-$\mathrm{T_{c}}$ systems such as doped $\mathrm{La_{2}CuO_{4}}$, where the orbital corresponds to a Cu $\mathrm{d_{x^2-y^2}}$ orbital. A related model on a 2D square lattice in the xy plane describes a ferromagnetic system in which the on-site states all have the same spin and the states $+$ and $-$ correspond to  $\mathrm{d_{z^2-x^2}}$ and  $\mathrm{d_{z^2-y^2}}$  $\mathrm{t_{2g}}$ orbitals respectively. This "$\mathrm{t_{2g}}$ model" describes ferromagnetic planes in  $\mathrm{Sr_{2}VO_{4}}$ and also in fluorides such as  $\mathrm{K_{2}CuF_{4}}$ and  $\mathrm{Cs_{2}AgF_{4}}$, where in this case a crystal field converts $\mathrm{e_g}$ orbitals into an effective $\mathrm{t_{2g}}$ 
   system~\cite{DWOAH08,WDOH08,MOKT05,HIYW83,MDTTBPSB06}. In the $\mathrm{t_{2g}}$ model one considers only the dominant hopping processes in which hopping between $+$~states occurs along the x axis and hopping between $-$~states occurs along the y axis. Thus in the $\mathrm{t_{2g}}$ model
 \begin{equation}\label{t2gHop}
  t_{\ro ab}=t_{0}\delta_{ab}\abs{\hat{\ro}\cdot\mathbf{e}_a}\,,
  \end{equation}
  where $\mathbf{e}_+$ and  $\mathbf{e}_-$ are unit vectors along the $x$ and $y$ directions respectively and $\hat{\ro}$  is a unit vector in the nearest-neighbour direction $\ro$.

In the large-$U$ limit we can approximate the Hamiltonian ${\cal H}_{\rm Hu}$ by an effective Hamiltonian ${\cal H}_{\rm eff}$ which is defined to act only within a subspace where there is no double occupation of any site. ${\cal H}_{\rm eff}$ takes the form
\begin{equation}\label{Heff}
{\cal H}_{\rm eff}={\cal T}- {\cal T}_1^2/U\,,
\end{equation}
where ${\cal T}_1$ is that part of ${\cal T}$ which has matrix elements linking states of no double occupation with states having a single doubly-occupied site. If three-site terms are neglected it is straight-forward to write the second term of Eq.~\eqref{Heff} in terms of spin (pseudospin for the $\mathrm{t_{2g}}$ model) operators $S_{\rr}^z=(n_{\rr +}-n_{\rr -})/2$, $S_{\rr}^\pm=S_{\rr}^x\pm S_{\rr}^y=c_{\rr \pm}^{\dagger}c_{\rr \mp}$ and number operator $n_{\rr}=n_{\rr +}+n_{\rr -}$. Then, for the Hubbard model, ${\cal H}_{\rm eff}$ becomes
\begin{eqnarray}\label{HtJ}
{\cal H}_{\rm t-J}&=&{\cal T}_{\rm t-J}+\tfrac{1}{2}J \sum_{\rr,\ro}[S_{\rr}^z S_{\rr+\ro}^z+\tfrac{1}{2}(S_{\rr}^+S_{\rr+\ro}^-+ S_{\rr}^-S_{\rr+\ro}^+)-\tfrac{1}{4}n_{\rr}n_{\rr+\ro}]\nonumber\\&=&{\cal T}_{\rm t-J}+\tfrac{1}{2}J \sum_{\rr,\ro}(\mathbf{S}_{\rr}\cdot\mathbf{S}_{\rr+\ro}-\tfrac{1}{4}n_{\rr}n_{\rr+\ro})\,,
\end{eqnarray}
where $J=4t_0^2/U$ and ${\cal T}_{\rm t-J}$ contains the hopping parameters of Eq.~\eqref{HubHop}~\cite{CSO77}.
For the $\mathrm{t_{2g}}$ model the corresponding Hamiltonian is
\begin{equation}\label{Ht2g}
{\cal H}_{\rm t_{2g}}={\cal T}_{\rm t_{2g}}+ \tfrac{1}{4}J \sum_{\rr,\ro}(S_{\rr}^z S_{\rr+\ro}^z-\tfrac{1}{4}n_{\rr}n_{\rr+\ro})\,,
\end{equation}
where ${\cal T}_{\rm t_{2g}}$ contains the hopping parameters of Eq.~\eqref{t2gHop}\cite{DWOAH08}.
${\cal H}_{\rm t-J}$ is known as the t-J model and if the transverse exchange terms in Eq.~\eqref{HtJ} are omitted we have the t-$\mathrm{J^z}$ model. This differs from ${\cal H}_{\rm t_{2g}}$ where $+$ and $-$ spins only hop along the x and y axes respectively. For the case of one electron per site, the undoped case for the oxides being modelled, ${\cal T}$ drops out and the system is a Mott insulator. In the Hubbard case
the insulator is a Heisenberg antiferromagnet and in the $\mathrm{t_{2g}}$ case it is an Ising antiferromagnet with alternating orbital order in the ground state. The transverse exchange terms in Eq.~\eqref{HtJ} are not present in Eq.~\eqref{Ht2g} owing to the directional hopping of the orbitals in the $\mathrm{t_{2g}}$ case. 

When a hole is introduced into the lattice its motion disturbs the spin or orbital order of the ground state. At each hop through the ordered lattice the hole leaves a spin, or orbital, deviation at the site it vacates. Thus as the hole hops through the lattice the energy of the system increases linearly and the hole is bound to its starting-point. This is known as the string effect and it exists in dimensions higher than one\cite{BR70}. In 1D the spin deviation created by a hole at its first hop, from some initial position, increases the energy but subsequent hops in the same direction merely shift the ordered spin configuration by a lattice spacing without further increase in energy. There is therefore no string effect in the 1D t-J or t-$\mathrm{J^z}$ models. Furthermore in the 2D or 3D t-J models the string effect is relaxed by the $S_{\rr}^+S_{\rr+\ro}^-$ terms in Eq.~\eqref{HtJ} which exchange spin directions between lattice sites and can lead to a healing of the spin deviations created by the hole. Thus the hole can move at a speed determined by the healing rate, which leads to a quasiparticle band of width proportional to the exchange parameter $J$~\cite{KLR89}. This way of relaxing the string effect is not possible in the t-$\mathrm{J^z}$ or the ${\cal H}_{\rm t_{2g}}$ model owing to the absence of the   
$S_{\rr}^+S_{\rr+\ro}^-$ terms. In these models a relaxation mechanism can be introduced by including in the Hamiltonian a term of the form
\begin{equation}\label{transfield}
 \sum_{\rr}(S_{\rr}^+ +S_{\rr}^-)=2\sum_{\rr}S_{\rr}^x
 \end{equation}
 so that the spin part of Eq.~\eqref{Ht2g} becomes a transverse-field Ising antiferromagnet. In the physical case of the $\mathrm{t_{2g}}$ model this transverse field corresponds to an on-site crystal field which mixes the two $\mathrm{t_{2g}}$ orbitals just as the transverse magnetic field mixes $+$ and $-$ spins. In both the t-$\mathrm{J^z}$ and the ${\cal H}_{\rm t_{2g}}$ model inclusion of three-site terms mentioned earlier also relaxes the string effect~\cite{DWOAH08} but we shall not include them here. A hole in the 2D t-$\mathrm{J^z}$ model can also propagate by means of a Trugman path~\cite{Tr88} which consists of 6 hops around a 4-site square plaquette. The hole moves to a next-nearest neighbour site and leaves the antiferromagnetic spin arrangement undisturbed. This cannot occur in the ${\cal H}_{\rm t_{2g}}$ model owing to the directional hopping~\cite{DWOAH08}.
We shall here concentrate on the t-J model and the t-$\mathrm{J^z}$ model in a transverse field although there should be no difficulty in treating the directional hopping of the ${\cal H}_{\rm t_{2g}}$ model. 

Our reference state has one electron on each site with spins (or pseudospins) ordered as in an antiferromagnet. We follow Martinez and Horsch~\cite{MH91a} in introducing a spinless fermion operator $f_{\rr}^\dagger$ which creates a hole in the reference state at site $\rr$ and a boson operator $b_{\rr}^\dagger$
which creates a spin reversal on site $\rr$. Thus $b_{\rr}^\dagger=S_{\rr}^ -$ for site $\rr$ on the $+$ spin sublattice and $b_{\rr}^\dagger=S_{\rr}^+$ for $\rr$ on the $-$ sublattice. Clearly each nearest-neighbour hop of the hole through the ordered lattice reverses the spin on the site vacated by the hole. Thus the hopping operator ${\cal T}_{\rm t-J}$ may be written as
\begin{equation}\label{ThtJ}
{\cal T}_{\rm t-J}=-t_0\sum_{\rr \ro}(f_{\rr+\ro}^\dagger f_{\rr}^{}b_{\rr}^\dagger+{\rm H.c.})\,.
\end{equation}
For the $\mathrm{t_{2g}}$ model the corresponding hopping operator ${\cal T}_{\rm t_{2g}}$ is
\begin{equation}\label{Th2g}
{\cal T}_{\rm t_{2g}}=-t_0\sum_{\rr \ro}(f_{\rr+\ro}^\dagger f_{\rr}^{}b_{\rr}^\dagger +{\rm H.c.})\,\abs{\hat{\ro}\cdot\mathbf{e}_a}\,.
\end{equation}
Since $S_{\rr}^z=\pm(\tfrac{1}{2}-b_{\rr}^\dagger b_{\rr})$ for $\rr$ on the $\pm$ spin sublattice we may write the Ising part of the exchange term, in the presence of a hole, as
\begin{equation}\label{Isingexch}
-\tfrac{1}{8}J\sum_{\rr \ro}(1-h_{\rr})(1-2b_{\rr}^\dagger b_{\rr}^{})(1-2b_{\rr+\ro}^\dagger b_{\rr+\ro})(1-h_{\rr+\ro})
=\tfrac{1}{2}J\sum_{\rr \ro}b_{\rr}^\dagger b_{\rr}^{}(1-h_{\rr+\ro}-b_{\rr+\ro}^\dagger b_{\rr+\ro}^{})\,,
\end{equation}
where $h_{\rr}=f_{\rr}^\dagger f_{\rr}^{}$ and an irrelevant constant term has been dropped. Here we have used $h_{\rr}b_{\rr}^\dagger b_{\rr}=0$ since the hole and a spin deviation cannot occupy the same site. A spin deviation created on site $\mathrm{\rr}$ when the hole vacates that site will in general have one neighbouring site  $\mathrm{\rr+\ro}$ occupied by the hole and another by a spin deviation created when the hole arrived on site $\mathrm{\rr}$. Thus the expression in Eq.~\eqref{Isingexch} may be written as 
\begin{equation}\label{Isingexch2}
\tfrac{1}{2}J(z-2)\sum_{\rr}b_{\rr}^\dagger b_{\rr}^{}\,,
\end{equation}
where $z$ is the number of nearest neighbours. $z=2$ in 1D so that the above term is zero, which is consistent with the absence of the string effect in the 1D t-$\mathrm{J^z}$ model. Apart from an additive constant the t-J Hamiltonian Eq.~\eqref{HtJ} becomes
\begin{equation}\label{Hh}
{\cal H}_{\rm t-J}=-t_0\sum_{\rr \ro}(f_{\rr+\ro}^\dagger f_{\rr}^{}b_{\rr}^\dagger +{\rm H.c.})+\tfrac{1}{4}J\sum_{\rr \ro}(b_{\rr}^\dagger b_{\rr+\ro}^\dagger +b_{\rr}^{}b_{\rr+\ro}^{})+\tfrac{1}{2}J(z-2)\sum_{\rr}b_{\rr}^\dagger b_{\rr}^{}\,.
\end{equation}
Note that the derivation of this Hamiltonian in Ref.~\cite{MH91a} 
gives additional constraints which we will not discuss here. 

The relaxation of the string effect occurs when bosons, created by fermion hopping, are spontaneously destroyed in nearest neighbour pairs by the terms $b_{\rr}b_{\rr+\ro}$. Edwards~\cite{Ed06} introduced a simplified model Hamiltonian of the form
\begin{equation}\label{EdwardsHam} 
 {\cal H}_{\rm Ed}=-t_0\sum_{\rr \ro}(f_{\rr+\ro}^\dagger f_{\rr^{}}^{}b_{\rr}^\dagger 
+ {\rm H.c.})-\lambda\sum_{\rr}(b_{\rr}^\dagger +b_{\rr}^{})+\omega_0\sum_{\rr}b_{\rr}^\dagger b_{\rr}^{}+N\lambda^2/\omega_0
\end{equation}
in which boson relaxation terms $b_{\rr}b_{\rr+\ro}$ are replaced by the simpler linear ones $b_{\rr}$. A unitary transformation $\tilde{\cal H}_{\rm Ed}={\rm e}^{\cal S} {\cal H}_{\rm Ed} {\rm e}^{-{\cal S}}$, ${\cal S}=(\lambda/\omega_0)\sum_{\rr}(b_{\rr}-b_{\rr}^\dagger)$, resulting in $b_{\rr}\rightarrow b_{\rr}+\lambda/\omega_0$, yields
\begin{equation}\label{EdwardsHamBar}
\tilde{\cal H}_{\rm Ed}=-\frac{2t_0 \lambda}{\omega_0}\sum_{\rr \ro}f_{\rr+\ro}^\dagger f_{\rr}^{}-t_0\sum_{\rr \ro}(f_{\rr+\ro}^\dagger f_{\rr}^{}b_{\rr}^\dagger + {\rm H.c.})+\omega_0\sum_{\rr}b_{\rr}^\dagger b_{\rr}^{}\,.
\end{equation}
Thus the second term in ${\cal H}_{\rm Ed}$ is eliminated in favour of the first term in $\tilde{\cal H}_{\rm Ed}$ which introduces a coherent hopping channel in addition to the original incoherent one. In the ground state $|\tilde{0}\rangle$  of the Hamiltonian $\tilde{\cal H}_{\rm Ed}$, in the absence of fermions, there are no bosons, so that $b_{\rr}|\tilde{0}\rangle=0$. Thus for the ground state $|0\rangle$ of ${\cal H}_{\rm Ed}$ we have $(b_{\rr}-\lambda/\omega_0)|0\rangle=0$ and hence $\langle0|b_{\rr}^\dagger b_{\rr}|0\rangle=\lambda^2/\omega_0^2$. 
 
 From the above discussion it is clear that in 2D the Edwards model corresponds, within a certain range of parameters, to an underlying t-$\mathrm{J^z}$ type of model with Hamiltonian
\begin{equation}\label{transfieldHam}
-t_0\sum_{\rr \ro \sigma}c_{\rr+\ro}^\dagger c_{\rr \sigma}^{}+\tfrac{1}{2}\omega_0\sum_{\rr \ro}S_{\rr}^z S_{\rr+\ro}^z -2\lambda\sum_{\rr}S_{\rr}^x\,,
\end{equation}
since $2S_{\rr}^x=S_{\rr}^+ +S_{\rr}^- =b_{\rr}^\dagger +b_{\rr}.$ This corresponds to a doped antiferromagnetic Ising model in a transverse magnetic field, but this correspondence is only valid for low hole density (low fermion density in the ${\cal H}_{\rm Ed}$ model) and $\lambda\ll\omega_0$. In this case, since $|\langle0|S_{\rr}^z|0\rangle|=\tfrac{1}{2}-\langle0|b_{\rr}^\dagger b_{\rr}|0\rangle =\tfrac{1}{2}-\lambda^2/\omega_0^2$, the holes move in a background medium which is close to a saturated antiferromagnet. For $\lambda\gtrsim\omega_0$ the transverse-field Ising model no longer exhibits antiferromagnetic order~\cite{Sti73}.
In 1D the correspondence breaks down completely since if $\lambda=0$ and $\omega_0\neq0$ the ${\cal H}_{\rm Ed}$ model exhibits the string effect which is absent in the 1D  t-$\mathrm{J^z}$ model. In fact the 1D ${\cal H}_{\rm Ed}$ model has an interesting similarity to the 2D t-$\mathrm{J^z}$ model. Even for $\lambda=0$ the string effect is relaxed by an analogue of the Trugman path~\cite{AEF07}. This is again a 6-step process in which the fermion propagates to a next-nearest neighbour site with the background medium left undisturbed, i.e. no bosons excited. During the process the fermion excites three bosons which are subsequently destroyed. Unfortunately the model ${\cal H}_{\rm Ed}$ does not represent precisely any physical system that we know of. The underlying Ising exchange is characteristic of the $\mathrm{t_{2g}}$ model for vanadates and fluorides but the directional hopping of that model is not included, although this could be remedied in the 2D model. However if the model is considered over the whole 
$\lambda/t_0$, $\omega_0/t_0$ space it is found, even in the 1D case, to exhibit a surprising number of different physical regimes reminiscent of some found in realistic strongly-correlated electron systems and strongly-coupled electron-phonon systems. This was first demonstrated for the case of a single fermion ($N_f=1$) at temperature T=0 by the method of variational Lanczos diagonalisation~\cite{AEF07}. A great advantage of the model is that the simple treatment of the background medium, in terms of local bosons, makes it possible to obtain essentially exact results in the thermodynamic limit, at least in the 1D case. More recently the half-filled case, with $N_f=N/2$ where $N$ is the number of lattice sites, has been investigated~\cite{WFAE08,EHF09,EF09b}. The ground-state phase diagram has been mapped out in the whole $\lambda-\omega_0$ plane, using a density matrix renormalisation group (DMRG) technique~\cite{Wh92,Wh93}. A quantum phase transition between a metallic Tomonaga-Luttinger liquid and an insulating charge density wave (CDW) was shown to exist.

It is desirable to complement these numerical results with some more analytical approaches. Nearly all analytical work on t-J-like or polaronic models is confined to the case of a single fermion. However very recently the projective renormalisation method~\cite{BHS02} has been applied to the half-filled 1D ${\cal H}_{\rm Ed}$ model in a study of CDW formation at T=0~\cite{SBF10}. Our initial analytical work on the ${\cal H}_{\rm Ed}$ model was confined to the one-fermion case in 1D at T=0 with the additional restriction $\lambda=0$~\cite{AEF10}. The latter restriction means that the fermion can only propagate as a coherent quasiparticle by the Trugman-like process discussed above. This requires the coexistence of at least three bosons during the particle's motion and in Ref.~\cite{AEF10} the one-fermion Green's function was calculated within a 3-boson approximation. The spectral functions agree well with exact numerical results for $\omega_0/t_0\gtrsim1$.  In this paper we extend the analytical approach for $\lambda=0$ to finite fermion density and to higher dimension, 2D and 3D. This is achieved within a 2-boson approximation and comparison with exact numerical results in 1D shows that the range of validity is thereby reduced to $\omega_0/t_0\gtrsim3$. No Trugman-like processes exist in the 2-boson approximation but they are unimportant to understand CDW formation at finite density. Relaxation of the string effect does not depend on them because bosons created by one fermion can be destroyed by other fermions. Near half-filling a CDW state is found, in good agreement with 1D numerical results, and the CDW transition temperature can be calculated formally. However, within the 2-boson approximation, the transition is mean-field-like, with no short-range order in the disordered state. This is clearly not realistic in 1D but finite-temperature calculations should certainly be relevant in 3D. In this paper we concentrate on the calculation of spectral functions in 1D at T=0 with various fermion densities and compare with some new DMRG results.

In Sec.~\ref{sec:limitingcases} we determine the one-fermion Green's function for $\lambda=0$ within the 2-boson approximation in the cases of one particle ($N_f=1$) and one hole ($N_f=N-1$). The method used is different from that previously employed for $N_f=1$~\cite{AEF10} and provides a wave-function as a byproduct. In Sec.~\ref{sec:decoupling} we study the hierarchy of equations of motion for the Green's function at arbitrary density and devise a decoupling which leads to the correct Green's function in the 2-boson approximation for $N_f=1$ and $N_f=N-1$. This 2-boson Green's function is valid for any bipartite lattice in 1D, 2D or 3D and allows for two distinct sublattices so that the CDW state can be investigated. In Sec.~\ref{sec:specCDW} spectral functions and ground-state
properties calculated from the 2-boson Green's function are compared with those calculated numerically by the DMRG method. In Sec.~\ref{conclusions} we draw conclusions and consider the outlook for future work.

\section{Limiting cases of the Green's function within the 2-boson approximation}\label{sec:limitingcases}
We now derive expressions for the one-fermion Green's function in the special cases of a single fermion ($N_f=1$) and a single hole ($N_f=N-1$). For these single-particle cases the Green's function may be calculated by direct solution of the Schr\"odinger equation, which also yields the wave-function. \subsection{The case $N_f=1$}
For $\lambda=0$ the Hamiltonian Eq.~\eqref{EdwardsHam}, written in $\mathrm{\kk}$-space, takes the form
\begin{equation}\label{hamNf=1}
{\cal H}_{\rm Ed}=\frac{1}{\sqrt{N}}\sum_{\kk \kk^\prime} f_{\kk}^\dagger f_{\kk^\prime} 
[\gamma(\kk)b_{\kk^\prime-\kk}^{\dagger} +\gamma(\kk^\prime)b_{\kk-\kk^\prime}]+\omega_0 \sum_{\qq} b_{\qq}^\dagger b_{\qq}^{}\,,
\end{equation}
where
\begin{equation}\label{FToperators}
f_{\kk}= \frac{1}{\sqrt{N}}\sum_{\rr} {\rm e}^{\ii\kk\cdot\rr} f_{\rr}\,,\qquad  
        b_{\qq}= \frac{1}{\sqrt{N}}\sum_{\rr} {\rm e}^{\ii\qq\cdot\rr} b_{\rr}\,,\qquad  \gamma(\kk)=-t_0\sum_{\ro} {\rm e}^{\ii\kk\cdot\ro}\,.
\end{equation}
The $\ro$-summation is over $z$ nearest neighbours in a 1D, 2D or 3D bipartite lattice. In the 2-boson approximation the wave-function for a single fermion is of the form
\begin{equation}\label{Psi}
\Psi_{\kk} =\Big[f_{\kk}^\dagger +\sum_{\qq_1} a(\qq_1)f_{\kk-\qq_1}^\dagger b_{\qq_1}^\dagger +\sum_{\qq_1} \sum_{\qq_2} a(\qq_1,\qq_2)f_{\kk-\qq_1-\qq_2}^\dagger b_{\qq_1}^\dagger b_{\qq_2}^\dagger \Big] |\mathrm{vac}\rangle\,,
\end{equation}
where $|\mathrm{vac}\rangle$ is the vacuum state and $a(\qq_1,\qq_2)=a(\qq_2,\qq_1)$. On substituting this in the Schr\"odinger equation ${\cal H}\Psi_{\kk }=E\Psi_{\kk}$, and multiplying on the left by $\langle\mathrm{vac}|$,
 $\langle\mathrm{vac}|b_{\qq^\prime} f_{\kk-\qq^\prime}$ and $\langle\mathrm{vac}|b_{\qq_1^\prime}  b_{\qq_2^\prime}f_{\kk-\qq_1^\prime-\qq_2^\prime}$ we obtain equations of the form
\begin{equation}\label{givesa}
-E+\frac{1}{\sqrt{N}}\sum_{\qq_1}\gamma(\kk-\qq_1)a(\qq_1)=0\,,
\end{equation}
\begin{equation}\label{givesaa}
\frac{1}{\sqrt{N}}\gamma(\kk-\qq_1)+a(\qq_1)(\omega_0 -E)+\frac{2}{\sqrt{N}}\sum_{\qq_2}
a(\qq_1,\qq_2)\gamma(\kk-\qq_1 -\qq_2)=0\,,
\end{equation}
\begin{equation}\label{givesaaa}
\frac{1}{\sqrt{N}}\gamma(\kk-\qq_1-\qq_2)[a(\qq_1)+a(\qq_2)]+2a(\qq_1,\qq_2)(2\omega_0 -E)=0\,.
\end{equation}
On solving Eq.~\eqref{givesaaa} for $a(\qq_1,\qq_2)$, and substituting in  Eq.~\eqref{givesaa}, we find
\begin{equation}\label{finala}
a(\qq_1)\left[\omega_0-E-\frac{zt_0^2}{2\omega_0 -E}\right]=-\frac{1}{\sqrt{N}}\gamma(\kk-\qq_1)+\frac{1}{N}
\sum_{\qq_2}\frac{[\gamma(\kk-\qq_1-\qq_2)]^2 a(\qq_2)}{2\omega_0 -E}\,.
\end{equation}
It is easily shown that for a bipartite lattice the last term in this equation vanishes when $a(\qq)\propto{\gamma(\kk-\qq)}$. Hence a solution for $a(\qq_1)$ is obtained by omitting the last term and, on substituting this solution in Eq.~\eqref{givesa}, we find
\begin{equation}\label{onefermionpoles}
E+\frac{zt_0^2}{\omega_0-E-\tfrac{zt_0^2}{2\omega_0 -E}}=0\,.
\end{equation}
The solutions of this equation are the energies of single-fermion eigenstates, which are the poles of the one-fermion Green's function $G_\kk(E)$. The left-hand side of Eq.~\eqref{onefermionpoles} is $G_\kk^{-1}(E)$. The absence of $\kk$-dependence in this expression shows that the single-fermion eigenstates are localized. This is due to the string effect which is not relaxed in the 2-boson approximation.
\subsection{The case $N_f=N-1$}
The motion of a single hole in the present model is quite different from that of a single particle. To avoid confusion it should be stressed that the hole discussed here does not correspond to a hole in a t-J-like model, the latter being represented by a fermion in the present model. Clearly when the hole considered here hops to a neighbouring site a boson is created on the arrival site, not on the departure site as in the motion of a single particle. This boson can be destroyed immediately when the hole makes a further hop. There is therefore no string effect and the hole propagates easily. To find the Green's function in this case we use the Schr\"odinger equation as in the case of a single particle. The wave-function for the hole is of the form
\begin{equation}\label{Phi}
\Phi_\kk =\Big[f_{\kk}+\sum_{\qq_1} c(\qq_1)f_{\kk+\qq_1}^{} b_{\qq_1}^\dagger +\sum_{\qq_1} \sum_{\qq_2} c(\qq_1,\qq_2)f_{\kk+\qq_1+\qq_2}^{} b_{\qq_1}^\dagger b_{\qq_2}^\dagger \Big] |F\rangle\,,
\end{equation}
where $|F\rangle$ is the state with every site occupied by a fermion and $c(\qq_1,\qq_2)=c(\qq_2,\qq_1)$. The equations corresponding to Eqs.~\eqref{givesa}-\eqref{givesaaa} in the previous case are
\begin{equation}\label{givesc} 
-E-\frac{1}{\sqrt{N}}\gamma(\kk)\sum_{\qq_1}c(\qq_1)=0\,,
\end{equation}
\begin{equation}\label{givescc}
-\frac{1}{\sqrt{N}}\gamma(\kk)+c(\qq_1)(\omega_0 -E)-\frac{2}{\sqrt{N}}\gamma(\kk+\qq_1)\sum_{\qq_2}
c(\qq_1,\qq_2)=0\,,
\end{equation}
\begin{equation}\label{givesccc}
-\frac{1}{\sqrt{N}}[c(\qq_1)\gamma(\kk+\qq_1)+c(\qq_2)\gamma(\kk+\qq_2)]+2c(\qq_1,\qq_2)(2\omega_0 -E)=0\,.
\end{equation}
From Eqs.~\eqref{givescc}-\eqref{givesccc} we find that for a bipartite lattice
\begin{equation}\label{finalc}
c(\qq_1)=\frac{1}{\sqrt{N}}\frac{\gamma(\kk)}{\omega_0-E-\tfrac{[\gamma(\kk+\qq_1)]^2}{2\omega_0 -E}}\,.
\end{equation}
Hence, from Eq.~\eqref{givesc},
\begin{equation}\label{oneholepoles}
-E-[\gamma(\kk)]^2 \frac{1}{N}\sum_\qq\frac{1}{\omega_0-E-\tfrac{[\gamma(\qq)]^2}{2\omega_0 -E}}=0\,.
\end{equation} 
The solutions of this equation are the energies of single-hole eigenstates and, as expected, they depend on the wave-vector $\kk$, being functions of 
$[\gamma(\kk)]^2$. This type of $\kk$ dependence arises from the fact that the hole propagates through the lattice, leaving behind no excited bosons, by means of double hops, as discussed at the beginning of this section. As in the one-particle case we deduce that for one hole ($N_f=N-1$) the one-fermion Green's function is given by
\begin{equation}\label{oneholeGreensfunction}
[G_\kk(E)]^{-1}=E-[\gamma(\kk)]^2 \frac{1}{N}\sum_\qq\frac{1}{\omega_0+E-\tfrac{[\gamma(\qq)]^2}{2\omega_0 +E}}\,.
\end{equation}
The sign of $E$ has been changed since in Eq.~\eqref{oneholepoles} the energy refers to a hole state.
\section{The Green's function at finite fermion density}\label{sec:decoupling}
In this section we study the hierarchy of equations of motion of the one-fermion Green's function and find a decoupling which is consistent with the results derived in Sec.~\ref{sec:limitingcases} for the limiting cases of low ($N_f=1$) and high ($N_f=N-1$) fermion density. We allow for two distinct sublattices with different occupation so that the CDW state can be investigated.

The Fourier transform of the one-fermion retarded Green's function is 
defined by~\cite{Zu60}
\begin{equation}\label{GkE}
G_\kk(E)=\langle\langle f_\kk;f_\kk^\dagger \rangle\rangle =-\ii \int_{-\infty}^{\infty}dt\theta(t)\langle[f_\kk(t),f_\kk]_+ \rangle {\rm e}^{\ii Et}\,,
\end{equation}
where $f_\kk(t)={\rm e}^{\ii Ht}f_\kk {\rm e}^{-\ii Ht}$ and $\theta(t)$ is the unit step function. We may write $f_\kk$, defined by Eq.~\eqref{FToperators}, as a sum of two sublattice components. Thus
\begin{equation}\label{fksub}
f_\kk=\frac{1}{\sqrt{2}}(f_{\kk 1}+f_{\kk 2})\,,
\end{equation}
where
\begin{equation}\label{FTsublat}
f_{\kk l}=\sqrt{\frac{2}{N}}\sum_{\rr_l}{\rm e}^{\ii\kk.\rr_l}f_{\rr_{l}}\,.
\end{equation}
Here the summation is over sites $\rr_l$ which belong to sublattice $l$ ($l=1,2$). It follows from Eqs.~\eqref{GkE} and ~\eqref{fksub} that
\begin{equation}\label{Gksub}
G_\kk(E)=\frac{1}{2}\sum_{l=1}^2 \sum_{m=1}^2 G_\kk^{lm}(E)\,,
\end{equation}
where
\begin{equation}\label{Gklm}
G_\kk^{lm}=\langle\langle f_{\kk l};f_{\kk m}^\dagger\rangle\rangle =\frac{2}{N}\sum_{\rr_l \ess_m} {\rm e}^{\ii \kk\cdot(\rr_l - \ess_m)}G_{\rr_l \ess_m}\,,
\end{equation}
and
\begin{equation}\label{Glm}
G_{\rr_l \ess_m}= \langle\langle f_{\rr_l};f_{\ess_m}^\dagger\rangle\rangle \,.
\end{equation}
The equation of motion for the Green's function $\langle\langle A;B\rangle\rangle $ is~\cite{Zu60}
\begin{equation}\label{eqmot}
E\langle\langle A;B\rangle\rangle =\langle [A,B]_+ 
\rangle+\langle\langle [A,{\cal H}];B\rangle\rangle  \,.
\end{equation}
Hence, noting that ${\cal H}$ is given by Eq.~\eqref{EdwardsHam} with $\lambda=0$, we find
\begin{equation}\label{eqmotG}
EG_{\rr_l \ess_m}=\delta_{\rr_l \ess_m}-t_0\sum_{\ro} H_{\rr_l +\ro,\ess_m}^{\rr_l} -t_0\sum_{\ro} I_{\rr_l +\ro,\ess_m}^{\rr_l +\ro}\,,
\end{equation}
where
\begin{equation}\label{GFH}
 H_{\rr_l +\ro,\ess_m}^{\rr_l}=\langle\langle f_{\rr_l +\ro}^{}b_{\rr_l};f_{\ess_m}^\dagger\rangle\rangle\,,
 \end{equation}
 \begin{equation}\label{GFI}
 I_{\rr_{\overline{l}},\ess_m}^{\rr_l +\ro}=\langle\langle f_{\rr_{\overline{l}}}^{}b_{\rr_l  +\ro}^\dagger;f_{\ess_m}^\dagger\rangle\rangle \,.
 \end{equation}
 Here $\rr_{\overline{l}}$ is a general site on the sublattice $\overline{l}$ which complements the sublattice $l$ ($\overline{l}=2,1$ for $l=1,2$ respectively). $\rr_l +\ro$ is a particular site on the sublattice $\overline{l}$ but it is necessary to define $I$ more generally in order to close the equations of motion. However we first calculate the second term on the right of Eq.~\eqref{eqmotG}.

The equation of motion for $H$ is
\begin{equation}\label{eqmotH}
\begin{split}
(E-\omega_0) H_{\rr_l +\ro,\ess_m}^{\rr_l}=\delta_{\rr_l +\ro,\ess_m}\langle b_{\rr_l}\rangle 
-t_0\sum_{\ro^{\prime}}\langle\langle f_{\rr_l +\ro}^{}f_{\rr_l +\ro^{\prime}}^\dagger f_{\rr_l}^{};f_{\ess_m}^\dagger\rangle\rangle\\
-t_0\sum_{\ro^{\prime}}\langle\langle f_{\rr_l +\ro+\ro^{\prime}}^{} (b_{\rr_l +\ro}^{}+
b_{\rr_l +\ro +\ro^{\prime}}^\dagger)b_{\rr_l}^{};f_{\ess_m}^\dagger\rangle\rangle \,.
\end{split}
\end{equation}
The last term of this equation is produced by a process in which a fermion hops from $\rr_l +\ro$ to $\rr_l +\ro +\ro^\prime$ with either the creation of a boson on the vacated site $\rr_l +\ro$ or destruction of a boson on the arrival site  $\rr_l +\ro +\ro^\prime$. The latter boson must have been created by another fermion so that the term involving $b_{\rr_l +\ro +\ro^{\prime}}^\dagger b_{\rr_l}$ corresponds to a dynamical interaction between fermions. This should only be included if we consider two-fermion interactions consistently which goes beyond the effective Hartree-Fock treatment introduced below. We therefore neglect this term. In the second term on the right of Eq.~\eqref{eqmotH} we retain only the part involving fermions on two sites, thus taking $\ro^\prime =\ro$, and then make a Hartree-Fock type of approximation. Thus the second term becomes
\begin{equation}\label{HFapprox}
-t_0\langle\langle (1-n_{\rr_l +\ro})f_{\rr_l}^{};f_{s_m}^\dagger\rangle\rangle \simeq-t_0(1-\langle n_{\rr_l +\ro}\rangle)G_{\rr_l,\ess_m}\,,
\end{equation}
where $n_{\rr_l}=f_{\rr_l}^\dagger f_{\rr_l}^{}$. Also we write the average fermion occupation for sites on sublattice $l$ as $n_l =\langle n_{\rr_l}\rangle$ so that $ \langle n_{\rr_l +\ro}\rangle =n_{\overline{l}}$. Following this discussion Eq.~\eqref{eqmotH} becomes
\begin{equation}\label{nexteqmotH}
(E-\omega_0) H_{\rr_l +\ro,\ess_m}^{\rr_l}=\delta_{\rr_l +\ro,\ess_m}\langle b_{\rr_l}\rangle
-t_0(1-n_{\overline{l}})G_{\rr_l,\ess_m}
-t_0\sum_{\ro^{\prime}}J_{\rr_l +\ro+\ro^\prime ,\ess_m}^{\rr_l +\ro, \rr_l}\,,
\end{equation}
where
\begin{equation}\label{GFJ}
J_{\rr_l +\ro+\ro^\prime ,\ess_m}^{\rr_l +\ro, \rr_l}=\langle\langle f_{\rr_l +\ro+\ro^\prime} b_{\rr_l +\ro} b_{\rr_l};f_{\ess_m}^\dagger\rangle\rangle \,.
\end{equation}
In the equation of motion for $J$ the terms arising from $[f_{\rr_l +\ro+\ro^\prime},{\cal H}]$ involve three boson operators and we neglect them. The terms arising from $[b_{\rr_l +\ro} b_{\rr_l},{\cal H}]$ may be treated in the Hartree-Fock-like way used to obtain Eq.~\eqref{HFapprox}. Hence
\begin{equation}\label{eqmotJ}
(E-2\omega_0)J_{\rr_l +\ro+\ro^\prime ,\ess_m}^{\rr_l +\ro, \rr_l} =\delta_{\rr_l +\ro+\ro^\prime ,\ess_m} \langle b_{\rr_l +\ro} b_{\rr_l}\rangle -t_0(1-n_l)H_{\rr_l +\ro,\ess_m}^{\rr_l}\,.
\end{equation}
We approximate
 $\langle b_{\rr_l +\ro} b_{\rr_l}\rangle $ by $b_l^{} b_{\overline{l}}$, where $b_l=\langle b_{\rr_l}\rangle$. Furthermore it is shown below that $b_l=0$ for the present model, with $\lambda=0$. Hence  $ \langle b_{\rr_{l}+\ro} b_{\rr_{l}}\rangle$ may be neglected. On substituting for $J$ in Eq.~\eqref{nexteqmotH}, using Eq.~\eqref{eqmotJ}, and summing over $\ro$, we find
\begin{equation}\label{finalH}
\sum_{\ro} H_{\rr_l +\ro,\ess_m}^{\rr_l}=C(E,n_l)\Big[b_l \sum_{\ro}  
\delta_{\rr_l +\ro,\ess_m} -zt_0(1-n_{\overline{l}})G_{\rr_l,\ess_m}\Big]\,,
\end{equation}
where
\begin{equation}\label{C}
C(E,n_l)=\frac{1}{E-\omega_0 -\tfrac{zt_0^2(1-n_l)}{E-2\omega_0}}\,.
\end{equation}

The second term on the right of Eq.~\eqref{eqmotG} has therefore been determined and to find the third term we must consider the equation of motion of the Green's function $I$ defined by Eq.~\eqref{GFI}. This takes the form
\begin{equation}\label{eqmotI}
\begin{split}
(E+\omega_0) I_{\rr_{\overline{l}},\ess_m}^{\rr_l +\ro}=\delta_{\rr_{\overline{l}},\ess_{m}} \langle b_{\rr_{l} +\ro}^{\dagger}\rangle
+t_0\sum_{\ro^{\prime}}\langle\langle f_{\rr_{\overline{l}}}^{}f_{\rr_{l} +\ro}^{\dagger} f_{\rr_l +\ro+\ro^{\prime}}^{};f_{\ess_m}^{\dagger}\rangle\rangle\\
-t_0\sum_{\ro^{\prime}}\langle\langle f_{\rr_{\overline{l}}^{}+\ro^{\prime}}(\delta_{\rr_{\overline{l}},\rr_{l} +\ro} + b_{\rr_{l} +\ro}^{\dagger} b_{\rr_{\overline{l}}}^{} +b_{\rr_{\overline{l}} +\ro^{\prime}}^{\dagger} b_{\rr_{l} +\ro}^{\dagger});f_{\ess_{m}}^{\dagger} \rangle\rangle \,.
\end{split}
\end{equation}
We now make similar approximations to those used to obtain Eq.~\eqref{nexteqmotH}. Thus we make an effective Hartree-Fock approximation to the second term on the right of Eq.~\eqref{eqmotI}, retaining it only when $\rr_{\overline{l}}=\rr_l +\ro$, and we neglect the $b_{\rr_{l} +\ro}^\dagger b_{\rr_{\overline{l}}}$ term in the third term. Hence
\begin{equation}\label{nexteqmotI}
(E+\omega_0) I_{\rr_{\overline{l}},\ess_m}^{\rr_l +\ro}=
\delta_{\rr_{\overline{l}},\ess_m}b_{\overline{l}}^* -t_0 n_{\lbar}\delta_{\rr_{\lbar},\rr_l +\ro}\sum_{\ro^{\prime}}G_{\rr_{\lbar} +\ro^{\prime}, \ess_m} -t_0 \sum_{\ro^{\prime}}
K_{\rr_{\lbar} +\ro^{\prime}, \ess_m}^{\rr_{\lbar} +\ro^{\prime},\rr_l +\ro}\,,
\end{equation}
where
\begin{equation}\label{K}
K_{\rr_{\lbar} +\ro^{\prime}, \ess_m}^{\rr_{\lbar} +\ro^{\prime},\rr_l +\ro}=
\langle\langle f_{\rr_{\lbar}+\ro^{\prime}}b_{\rr_{\lbar}+\ro^{\prime}}^{\dagger} b_{\rr_{l}+\ro}^{\dagger};
f_{\ess_m}^{\dagger}\rangle\rangle \,.
\end{equation}
We treat the equation of motion of $K$ in a similar way to that of $J$. A slight difference is that instead of neglecting all the terms arising from $[f_{\rr_{\lbar}+\ro^{\prime}},H]$ we retain one which leads to a factor $b_{\rr_{\lbar} +\ro^{\prime}}b_{\rr_{\lbar} +\ro^{\prime}}^{\dagger}$. This equals $[1+b_{\rr_{\lbar} +\ro^{\prime}}^{\dagger}b_{\rr_{\lbar} +\ro^{\prime}}]$ which we approximate by 1, neglecting the boson occupation number. Hence, neglecting a $ \langle b^{\dagger}b^{\dagger}\rangle$ correlation function as before, we find
\begin{equation}\label{eqmotK}
(E+2\omega_0)K_{\rr_{\lbar} +\ro^{\prime}, \ess_m}^{\rr_{\lbar} +\ro^{\prime},\rr_l +\ro}=-t_0 n_l \sum_{\ro^{\prime\prime}} I_{\rr_{\lbar} +\ro^{\prime} +\ro^{\prime\prime},\ess_m}
^{\rr_l +\ro}\,.
\end{equation}
Thus, from Eqs.~\eqref{nexteqmotI} and ~\eqref{eqmotK}, 
\begin{equation}\label{moreeqmotI}
(E+\omega_0) I_{\rr_{\overline{l}},\ess_m}^{\rr_l +\ro}=
\delta_{\rr_{\overline{l}},\ess_m}b_{\overline{l}}^* -t_0 n_{\lbar}\delta_{\rr_{\lbar},\rr_l +\ro}\sum_{\ro^{\prime}}G_{\rr_{\lbar} +\ro^{\prime}, \ess_m} +\frac{t_0^2 n_l}{E+2\omega_0}\sum_{\ro^{\prime}}\sum_{\ro^{\prime\prime}}I_{\rr_{\overline{l}+\ro^{\prime}+ \ro^{\prime\prime}},\ess_m}^{\rr_l +\ro}\,.
\end{equation}
To solve this equation for $I$ we introduce the Fourier transform
\begin{equation}\label{FTI}
I_{\kk\qq}^{\lbar m}=\left(\frac{2}{N}\right)^{\frac{3}{2}}\sum_{\rr_{l}\rr_{\lbar}\ess_m} I_{\rr_{\overline{l}},\ess_m}^{\rr_l +\ro}\,\mathrm{e}^{\ii[(\kk+\qq)\cdot\rr_{\lbar}- \kk\cdot\ess_m-\qq\cdot(\rr_l +\ro)]}
\end{equation}
so that
\begin{equation}\label{finaleqmotI}
\Big(E+\omega_0-\tfrac{n_l [\gamma(\kk+\qq)]^2}{E+2\omega_0}\Big)
I_{\kk\qq}^{\lbar m}=\sqrt{\frac{N}{2}}\delta_{\qq 0}\delta_{\lbar m} b_{\lbar}^{*}+\sqrt{\frac{2}{N}}n_{\lbar} \gamma(\kk)G_{\kk}^{lm}\,.
\end{equation}
By taking the Fourier transform of Eq.~\eqref{eqmotG}, and using Eq.~\eqref{finalH}, we find
\begin{equation}\label{FTGlm}
EG_{\kk}^{lm}=\delta_{lm}-C(E,n_l)[b_l \gamma(\kk)\delta_{\lbar m}-z t_0^2 (1-n_{\lbar})G_{\kk}^{lm}]+\gamma(\kk)
I_{\kk}^{\lbar m}\,,
\end{equation}
where
\begin{equation}\label{moreFTI}
I_{\kk}^{\lbar m}=\frac{2}{N}\sum_{\rr_{\lbar} \ess_{m}} {\rm e}^{[\ii \kk\cdot (\rr_{\lbar}-\ess_m)]}I_{\rr_{\lbar} \ess_m}^{\rr_{\lbar}}=\sqrt{\frac{2}{N}}\sum_{\qq}I_{\kk \qq}^{\lbar m}\,.
\end{equation}
By combining the last three equations we obtain
\begin{equation}\label{Gklmnotfinal}
G_{\kk}^{lm}\big[E-zt_0^2(1-n_{\lbar})C(E,n_l)-[\gamma(\kk)]^2 n_{\lbar} \overline{D}(E,n_l)\big]
=\delta_{lm}-\big[C(E,n_l)b_l+D(E,\kk,n_l)b_{\lbar}^{*}\big]\delta_{\lbar m}\gamma(\kk)\,,
\end{equation}
where
\begin{equation}\label{D}
D(E,\kk,n_l)=\frac{1}{E+\omega_0-\frac{n_l[\gamma(\kk)]^2}{E+2\omega_0}}\,,
\end{equation}
and
\begin{equation}\label{Dbar}
\overline{D}(E,n_l)=\frac{2}{N}\sum_{\qq}D(E,\qq,n_l)\,.
\end{equation}
It remains to explain why, as indicated following Eq.~\eqref{eqmotJ}, $b_l=0$. This may be shown using
\begin{equation}\label{bl}
\begin{split}
b_l=\langle b_{\rr_{l}}\rangle & =\frac{1}{z}\sum_{\ro}\langle [b_{\rr_{l}}^{}f_{\rr_{l}+\ro}^{},f_{\rr_{l}+\ro}^{\dagger}]_{+}\rangle\\
& =\frac{\ii}{2\pi z}\sum_{\ro} \int_{-\infty}^{\infty}[H_{\rr_{l}+\ro,\rr_{l}+\ro}^{\rr_l}(E+\ii\eta) -H_{\rr_{l}+\ro,\rr_{l}+\ro}^{\rr_l}(E-\ii\eta)]dE\,.
\end{split}
\end{equation}
Eqs.~\eqref{finalH} and ~\eqref{Gklmnotfinal} may be used to show that the above expression is a linear combination of $b_l$ and $b_{\lbar}^{*}$. Thus $b_l, b_l^{*}\;(l=1,2)$ satisfy a set of linear homogeneous equations so that in general $b_l=0$. This result is consistent with the symmetry of the transverse-field t-$\mathrm{J^z}$ model (Eq.~\eqref{transfieldHam}) which underlies the present model for low fermion density. When $\lambda=0$, as assumed here, the transverse magnetic field vanishes so that by symmetry the expected value of the transverse spin moment $\langle S_{\rr}^x\rangle=\langle b_{\rr}^{\dagger}+b_{\rr}\rangle /2 =0$. The properties $b_l=0$ and the characteristic periodicity in ${\bf k}$  are shown in the Appendix  to be generally true  for the model with $\lambda=0$ for all values of $\omega_0/t_0$. As soon as $\lambda\neq0$ we shall in general have $b_l\neq0$ so that, from Eq.~\eqref{Gklmnotfinal}, the $\kk$ dependence of the Green's function will involve $\gamma(\kk)$ and not only $[\gamma(\kk)]^2$ as is the case for $\lambda=0$.

The final result for the Green's function in the case $\lambda=0$ is 
\begin{equation}\label{Gkmfinal}
G_{\kk}^{lm}(E)=\delta_{lm}[E-zt_0^2(1-n_{\lbar})C(E,n_l)-\gamma(\kk)^2 n_{\lbar} \overline{D}(E,n_l)]^{-1}
\end{equation}
where $C$ and $\overline{D}$ are given by Eqs.~\eqref{C} and~\eqref{Dbar} respectively. Also, from Eq.~\eqref{Gksub},
\begin{equation}\label{Gksubagain}
G_{\kk}(E)=\tfrac{1}{2}[G_{\kk 1}(E)+G_{\kk 2}(E)]\,,
\end{equation}
where $G_{\kk l}=G_{\kk}^{ll}$. In the homogeneous case, where there is no CDW, $n_1=n_2=n$, where $n=N_f /N$ is the fermion density, and $G_{\kk 1}=G_{\kk 2}=G_{\kk}$. For $n=0$ we recover the $N_f =1$ result for $G_{\kk}^{-1}$, given by the left-hand side of Eq.~\eqref{onefermionpoles}, and for $n=1$ we recover the $N_f =N-1$ result given by Eq.~\eqref{oneholeGreensfunction}.
Thus $G_{\kk}$ is correct in these two limits within the 2-boson approximation.
The factor $\delta_{lm}$ in Eq.~\eqref{Gkmfinal} shows that within the present approximation fermions propagate, by means of double hops, within a single sublattice. The dimension of the bipartite lattice (1, 2 or 3) enters only through the band energy $\gamma(\kk)$ [Eq.~\eqref{FToperators}] and the function $\overline{D}$, which may be written as
\begin{equation}\label{dbarint}
\overline{D}(E,n_l)=\int\frac{N_0(\gamma)}{E+\omega_0 -\gamma^2 n_l/(E+2\omega_0)}d\gamma\,,
\end{equation}
where $N_0(\gamma)$ is the density of states per site for the band energy $\gamma(\kk)$.

To complete the present formulation we require the equations which determine the chemical potential $\mu$ and the self-consistent sublattice densities $n_1, n_2$ for a given fermion density $n$. The density of states per site on sublattice $l$ is given by
\begin{equation}\label{sublatticedensityofstates}
N_l (E)=-\frac{1}{N\pi}\sum_{\kk}\Im G_{\kk l}(E+\ii\eta)
\end{equation}
and
\begin{equation}\label{sublattocc}
n_l=\int N_{l}(E)f(E,\mu)dE\,,
\end{equation}
where $f(E,\mu)=[{\rm e}^{\beta(E-\mu)}+1]^{-1}$ is the Fermi function with chemical potential $\mu$ and $\beta=(k_B T)^{-1}$.
The chemical potential and the CDW order parameter $P$ are determined by
\begin{equation}\label{nP}
\tfrac{1}{2}(n_1+n_2)=n,\quad\quad \tfrac{1}{2}(n_1-n_2)=nP\,.
\end{equation} 
If the system is ordered ($P\neq0$) at $T=0$ the CDW temperature $T_c$, where $P\rightarrow0$, can be calculated. In the present mean-field-like approximation this will not be the true $T_c$ but a higher temperature where short-range order substantially disappears. The calculated $T_c$ should be a reasonable approximation to the true value in 3D. In general, if quantities such as the spectral functions discussed below are calculated assuming $P=0$, the results will be valid in the high temperature limit where there is truly no short-range order.

The spectral function for states projected onto the sublattice $l$ is given by
\begin{equation}\label{projectedspectralfunction}
S_l(\kk,E)=-\frac{1}{\pi}\Im G_{\kk l}(E+\ii \eta)
\end{equation}
and the total spectral function is 
\begin{equation}\label{totalspectralfunction}
S(\kk,E)=-\frac{1}{\pi}\Im G_{\kk}(E+\ii \eta)=\frac{1}{2}[S_1(\kk,E)+S_2(\kk,E)].
\end{equation}
These spectral functions may be used to calculate the occupation number $n(\kk)=\langle f_\kk^{\dagger}f_\kk^{} \rangle$ and the related quantity $d(\kk)=\langle f_\kk^{\dagger}f_{\kk+\mathbf{Q}/2}^{} \rangle$, where $\mathbf{Q}$ is a basis vector of the reciprocal lattice. $f_\kk$ is given by Eq.~\eqref{fksub} and it follows from Eq.~\eqref{FTsublat} that $f_{\kk+\mathbf{Q}/2}=(f_{\kk 1}-f_{\kk 2})/\sqrt{2}$. Hence we find
\begin{equation}\label{nk}
n(\kk)=\frac{1}{2}\int^{\mu}_{-\infty}[S_1(\kk,E)+S_2(\kk,E)]dE
\end{equation}
and
\begin{equation}\label{dk}
d(\kk)=\frac{1}{2}\int^{\mu}_{-\infty}[S_1(\kk,E)-S_2(\kk,E)]dE\,.
\end{equation}
These quantities satisfy the sum rules 
\begin{equation}\label{srules}
\frac{1}{N}\sum_{\kk}n(\kk)=n\,,\quad \frac{1}{N}\sum_{\kk}d(\kk)=nP\,.
\end{equation} 

In the next section we report numerical results for all the above quantities, based on the approximate Green's function of Eq.~\eqref{Gkmfinal}, in the 1D case. Many of the results are compared with those of the DMRG method, both to assess the validity of the present approximation and sometimes to throw new light on the DMRG results. The application of the DMRG method to the present model has been described previously~\cite{EHF09,EF09b}.

\section{Numerical results for ground-state and spectral properties
 in 1D}\label{sec:specCDW}
In 1D the band energy $\gamma(\kk)=-2\cos{k}$, with the lattice constant taken as 1 and the unit of energy taken as $t_0$. The quantity $\overline{D}$ which appears in the Green's function (Eq.~\eqref{Gkmfinal}) can then be evaluated analytically using Eqs.~\eqref{D} and \eqref{Dbar}. The $\qq$-summation is most conveniently performed as a contour integral around the unit circle. The result is
\begin{equation}\label{DbaroneD}
\overline{D}(E,n_l)=D_{lr}\theta(a_l ^2-b_l ^2)-\ii D_{li}\theta(b_l ^2-a_l ^2)\,,
\end{equation}
where
\begin{equation}\label{DlrDli}
D_{lr}=\frac{1}{a_l \sqrt{1-b_l ^2 /a_l ^2}}\; ,\qquad D_{li}=\frac{1}{\sqrt{b_l ^2-a_l ^2}}
\end{equation}
with 
\begin{equation}\label{albl}
a_l =E+\omega_0 -\frac{2n_l}{E+2\omega_0}\; , \qquad b_l =-\frac{2n_l}{E+2\omega_0}\,.
\end{equation}
The density of states, given by Eq.~\eqref{sublatticedensityofstates}, can be evaluated similarly with the result
\begin{equation}\label{Nl}
N_l (E)=N_{l1}(E)+N_{l2}(E)\,,
\end{equation}
where
\begin{equation}\label{Nlone}
N_{l1} (E)=\frac{1}{\pi}\theta (a_l ^2-b_l ^2)\left[\frac{\theta(v_l ^2-u_l ^2)}{\sqrt{v_l ^2-u_l ^2}}+\theta(u_l ^2-v_l ^2)\left|\Im\left(\frac{1}{ \sqrt{u_l(E+\ii \eta)^2 -v_l(E+\ii\eta)^2}} \right)\right|\right]\,,
\end{equation}
\begin{equation}\label{Nltwo}
N_{l2} (E)=\frac{\theta(b_l ^2-a_l ^2)}{2\pi n_{\lbar}D_{li}} \sqrt{\frac{\sqrt{V_l ^2 +4}-\abs{V_l}}{2\abs{V_l}(V_l ^2 +4)}}
\end{equation}
with
\begin{eqnarray}\label{uvV}
u_l(E)& =&E-2(1-n_{\lbar})C(E,n_l)-2n_{\lbar}D_{lr}(E)\,,\\
v_l(E)& =&-2n_{\lbar}D_{lr}(E)\,,\\
V_l(E)& =&-\frac{E-2(1-n_{\lbar})C(E,n_l)}{2n_{\lbar}D_{li}(E)}\,.
\end{eqnarray}
The results presented below are for the case $\omega_0 =3$. By comparison with the DMRG results for the ground state properties $n(k)$ and $d(k)$, and with the dynamical DMRG~\cite{Je02b} (DDMRG) results for $S(k,E)$, it is found that this boson energy is large enough for many results of the present analytic approximation to be quite accurate.

\subsection{High and low fermion density}\label{hi_low}

\begin{figure}[h]
 \begin{minipage}{0.5\hsize}
  \centering
  \includegraphics[width=0.85\textwidth]{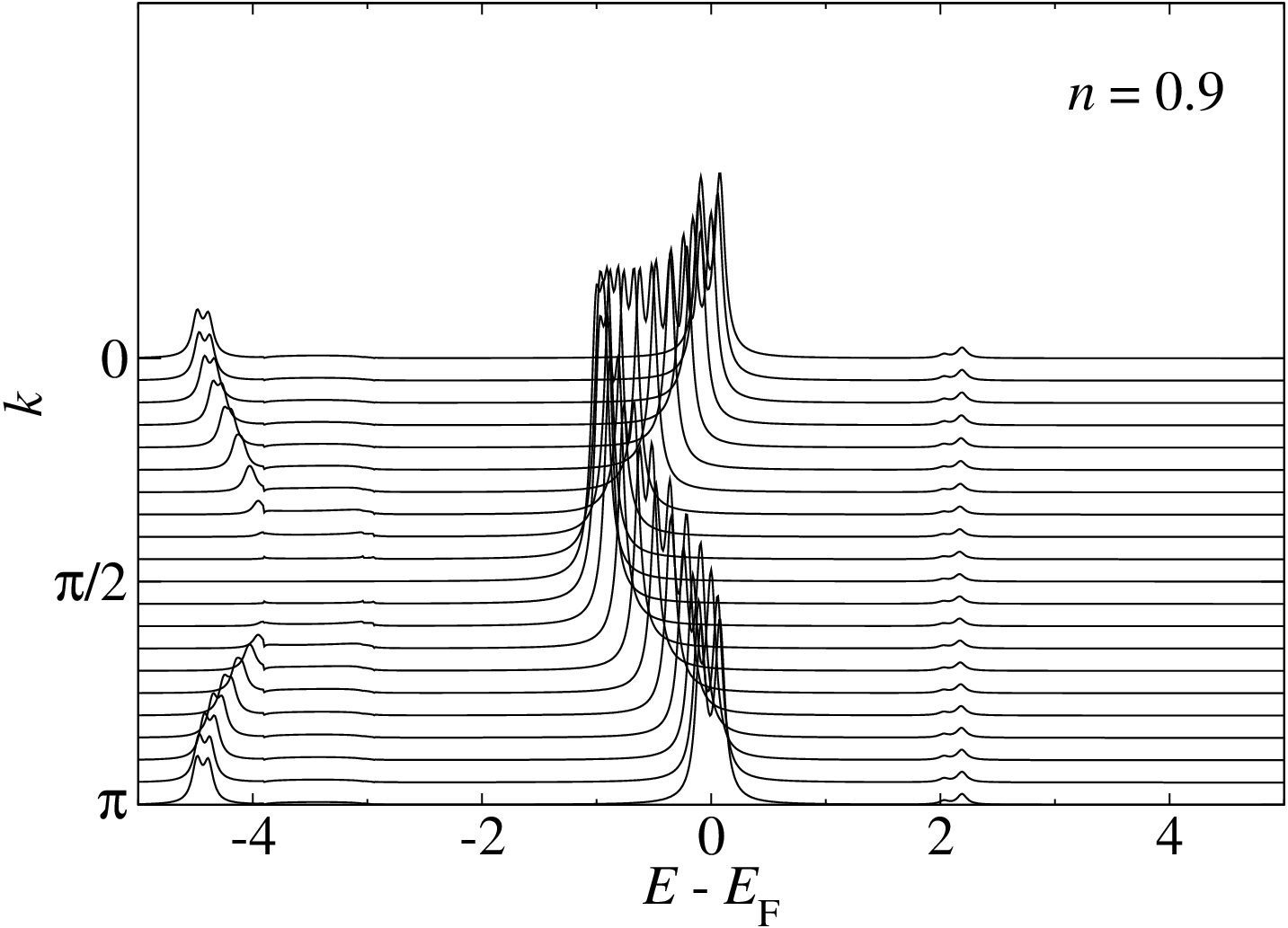}
 \end{minipage}
\begin{minipage}{0.5\hsize}
 \centering
 \includegraphics[width=0.85\textwidth]{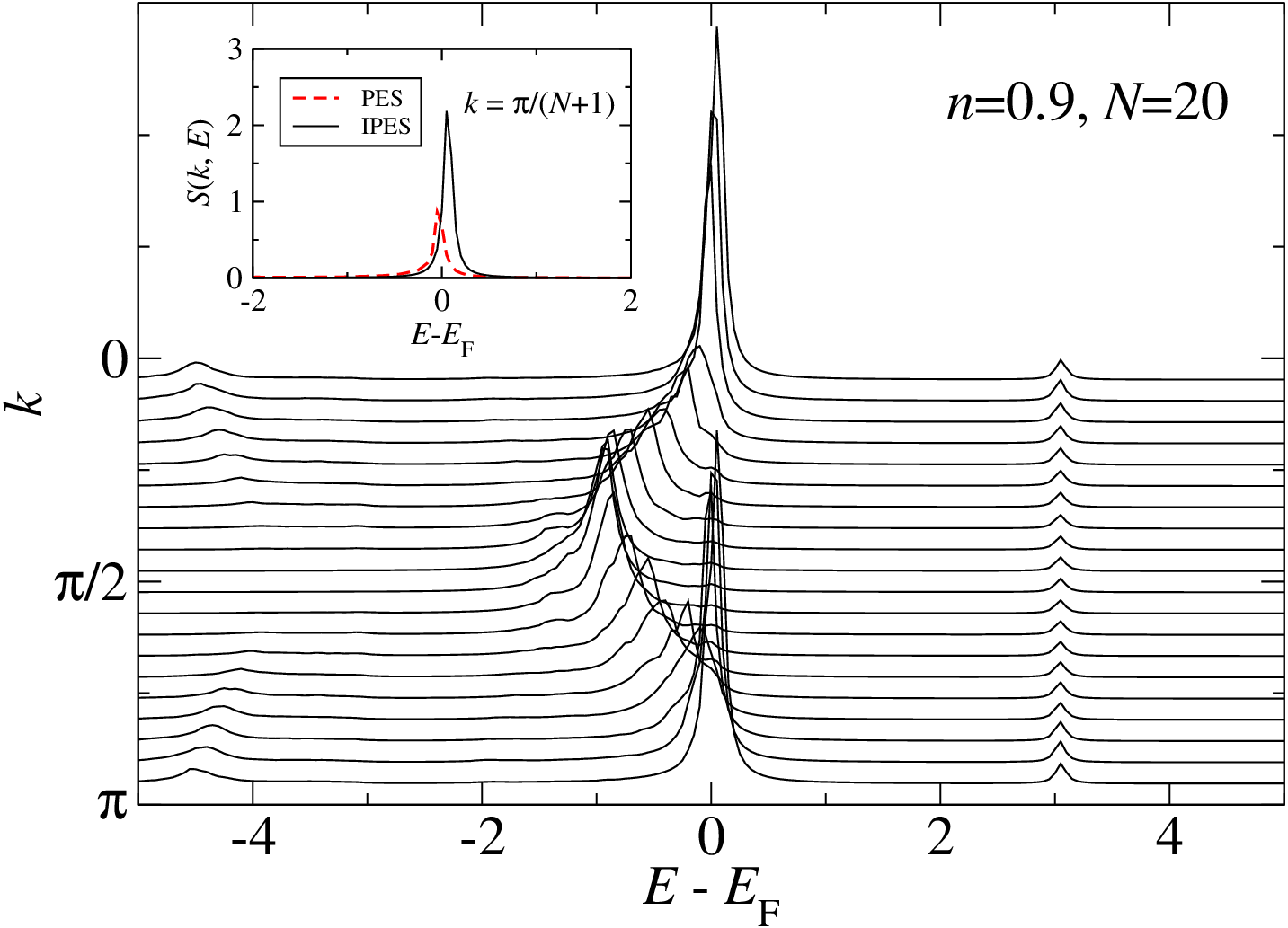}
\end{minipage} 
 \caption{Zero-temperature single-particle spectral function 
$S(k,E)$ at $n=0.9$ for $\omega_0=3$ from Eq.~(\ref{totalspectralfunction}) 
(left panel) compared with the DDMRG result (right panel) obtained 
for a finite system with $N=20$ sites using open boundary conditions (OBC). 
The inset shows the photoemission spectra (PES) and 
inverse photoemission spectra (IPES) near the Fermi point. 
$E$ is measured with respect to the Fermi 
energy $E_{\rm F}$ (all energies are given in units of $t_0$).
 }
\label{fig_spec_n09}
\end{figure}

The existence of a two-sub-lattice CDW state in the present model is well-established at half-filling ($n=0.5$) for sufficiently large 
$\omega_0/t_0$~\cite{WFAE08,EHF09,EF09b,SBF10}. However the general results of the Appendix  imply that a two-sublattice CDW state is always a possibility, whatever the density $n$. We therefore used Eqs.~(\ref{sublattocc}), (\ref{nP}) and~(\ref{Nl}) to search for such states even in the high and low density cases of $n=0.9$ and 0.1. In the case $n=0.9$ with $\omega_0/t_0 =3$ we find a self-consistent CDW state with order parameter $P=0.063$. The flow of the iterative procedure to determine $P$ indicates that this, not the uniform density $P=0$ state, is the stable ground state. In the DMRG calculations convergence to the CDW solution is improved by use of suitable external fields at the boundaries with open boundary conditions. The spectral function $S(k,E)$ calculated from the Green's function using Eq.~(\ref{totalspectralfunction}) is compared with the DDMRG results in Fig.~\ref{fig_spec_n09}. 
The quasiparticle peaks for the Green's function decoupling scheme results (left panels) are delta-functions in the limit $\eta\to 0$, but to make them visible we have taken $\eta =0.05$; the same value has been taken in the DDMRG 
data (right panel). The following main features are in good agreement: the general shape and width of the quasiparticle band crossing the Fermi level, the dispersive peaks just below $E-E_F =-4$ which vanish for $k=\pi/2$, the weak flat band at $E-E_F =2$ in the left panel and 3 in the right one. 
The splitting of the quasiparticle peaks due to CDW order is clearly visible in the left panel. This splitting is not clearly resolved in the main right panel but the inset for $k=\pi /21$ shows a splitting between a peak below $E_F$ in the photoemission spectrum (PES) and one above $E_F$ in the inverse photoemission spectrum (IPES).
%
%
The absence of the splitting for states further from $E_F$ in the DDMRG data is presumably due to finite lifetime broadening processes which are not included in the Green's function approximation. It should be noticed that the quasiparticle states at the Fermi level are from only one of the split subbands near $k=0$ and $\pi$, the other subband being fully occupied. This means that the Fermi wavevectors are at $\pi/10, \pi -\pi/10$ so that the hole pockets near $k=0$ and $\pi$ have the correct Fermi surface 'volume', this being twice what it would have been in the absence of the CDW ($P=0$). Discontinuities at these wavevectors are clearly seen in Fig. 3 where $n(k)$ and $d(k)$, calculated from Eqs.~(\ref{nk}) and~(\ref{dk}), are plotted.  
The agreement with the DMRG data, also plotted, is excellent. In the Green's function calculations the discontinuities are not perfectly sharp owing to the use of a finite $\eta$ (=0.0005) near $E_F$. There are two contributions to $n(k)$ and $d(k)$, one arising from the coherent quasiparticle bands and the other from spectral weight further below the Fermi level. In the present case $d(k)$ arises almost entirely from the coherent contribution and the quasiparticle states have slightly more weight on the minority sublattice, hence the negative value of $d(k)$ over most of the zone. Near $k=0$ and $\pi$, however, only the subband associated with the majority sublattice is occupied, hence the strong positive contribution. Correlations giving rise to Luttinger-liquid behaviour in 1D are beyond the present Green's function approach. It should be noted that at $n=0.9$ the CDW state is metallic, whereas at $n=0.5$ it is an insulator, as discussed in Sec.~\ref{CDW}.
\begin{figure}[b]
  \centering
  \includegraphics[width=0.48\textwidth]{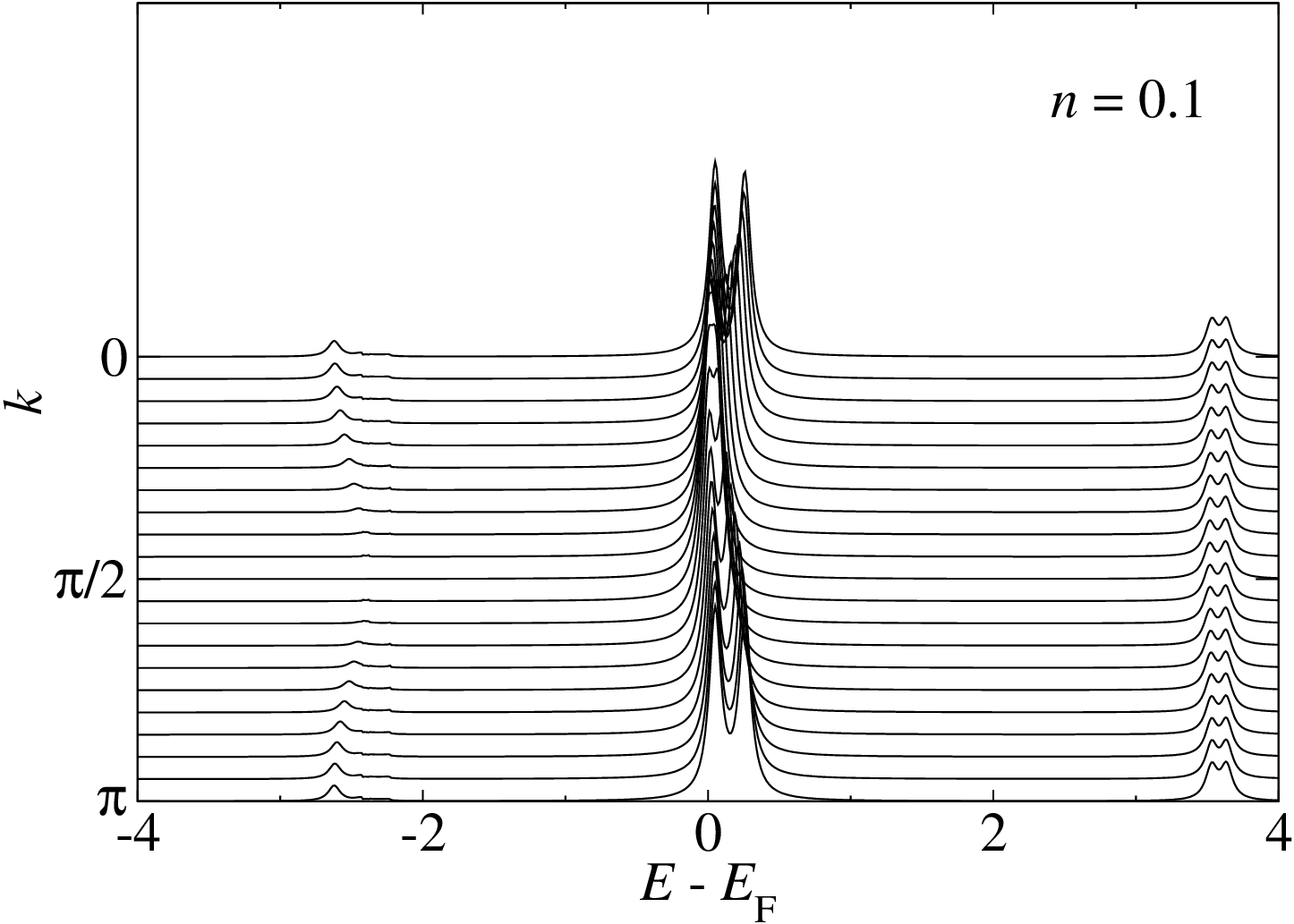}
\caption{Single-particle spectral function $S(k,E)$ at $n=0.1$
 for $\lambda=0$ and $\omega_0=3$ from Eq.~(\ref{totalspectralfunction}).} 
\label{fig_spec_n01}
\end{figure}

\begin{figure}[b]
 \begin{minipage}{0.5\hsize}
  \centering
  \includegraphics[width=0.85\textwidth]{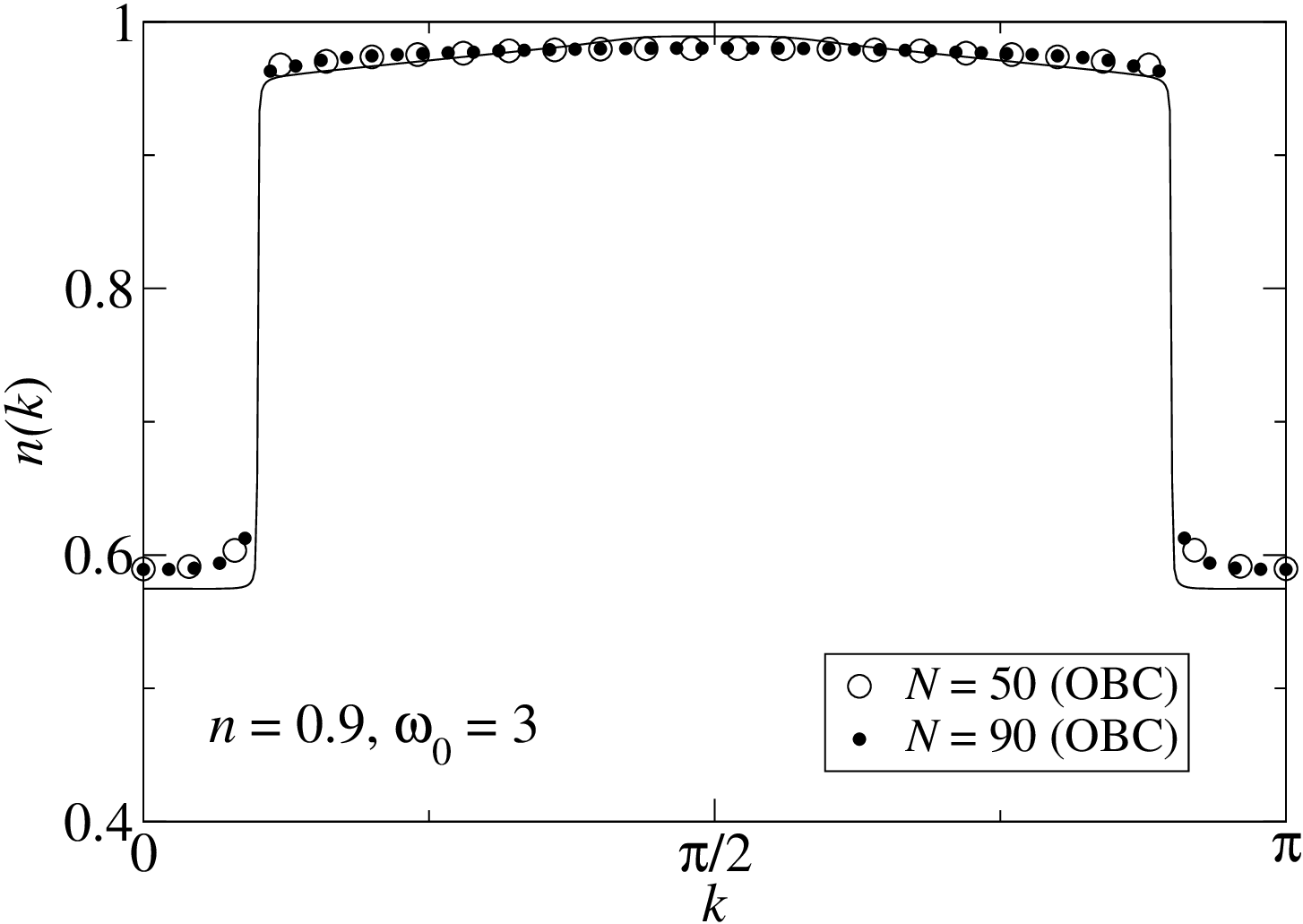}
 \end{minipage}
\begin{minipage}{0.5\hsize}
 \centering
 \includegraphics[width=0.85\textwidth]{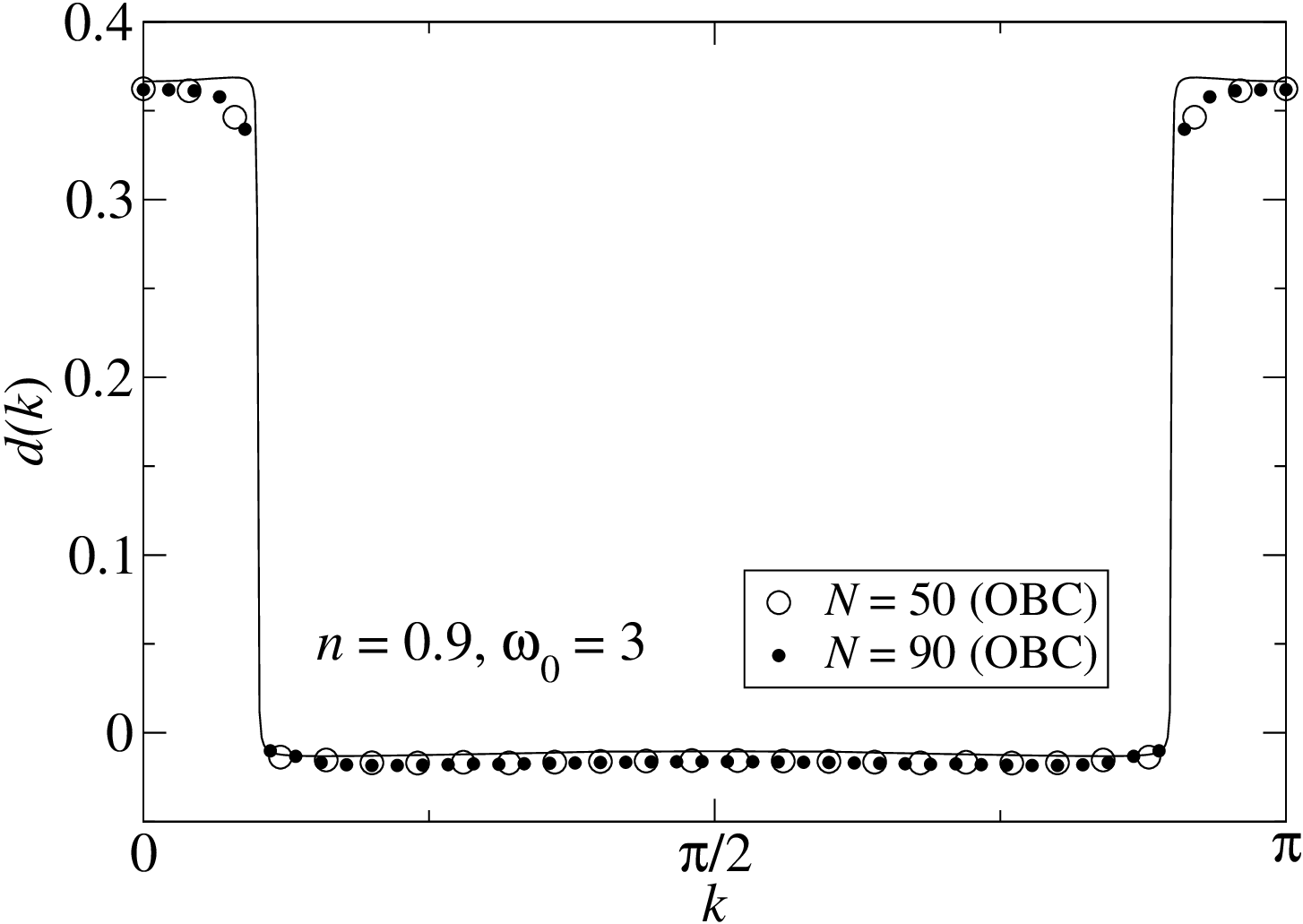}
\end{minipage}\\ 
\begin{minipage}{0.5\hsize}
  \centering
  \includegraphics[width=0.85\textwidth]{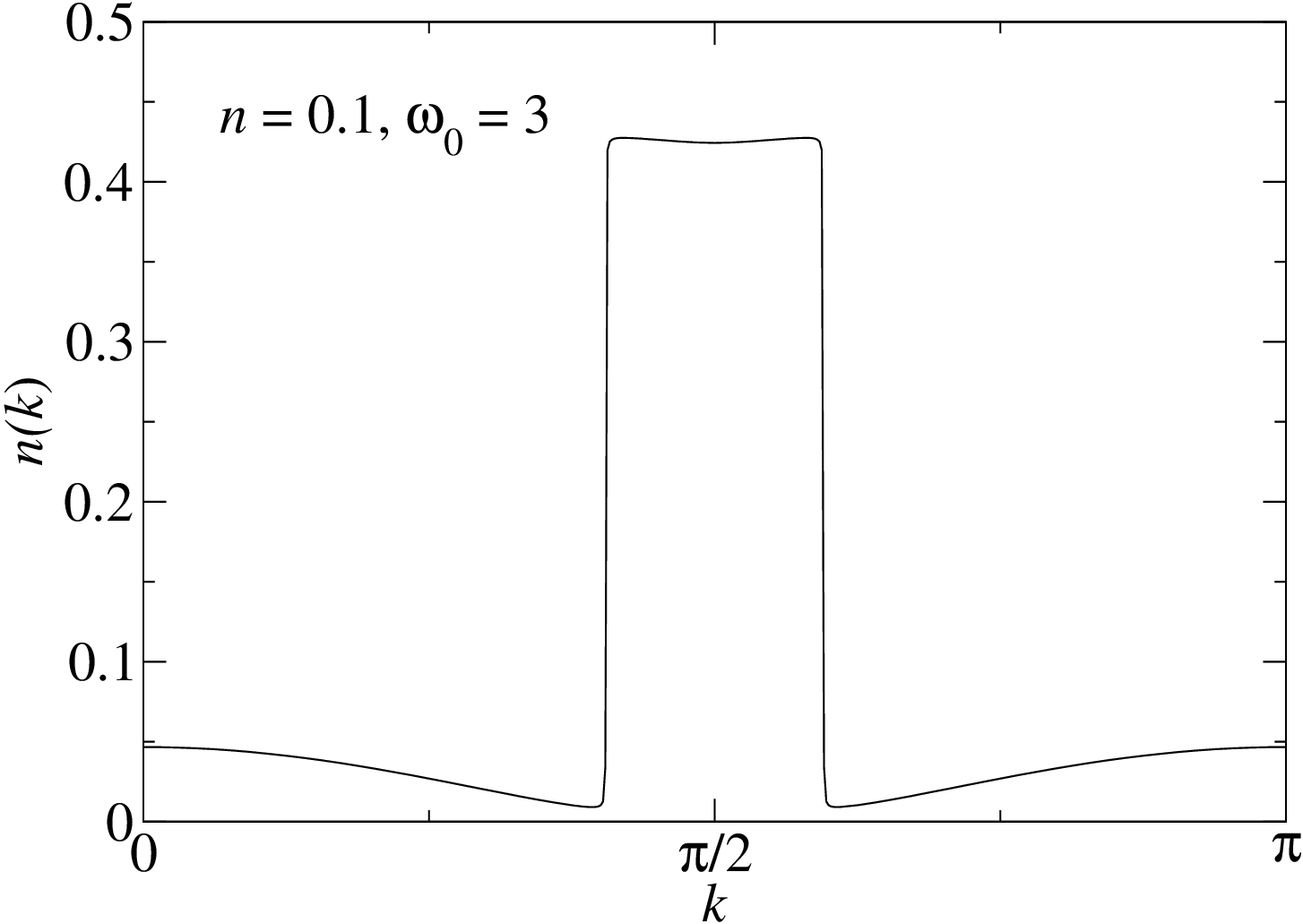}
 \end{minipage}
\begin{minipage}{0.5\hsize}
 \centering
 \includegraphics[width=0.85\textwidth]{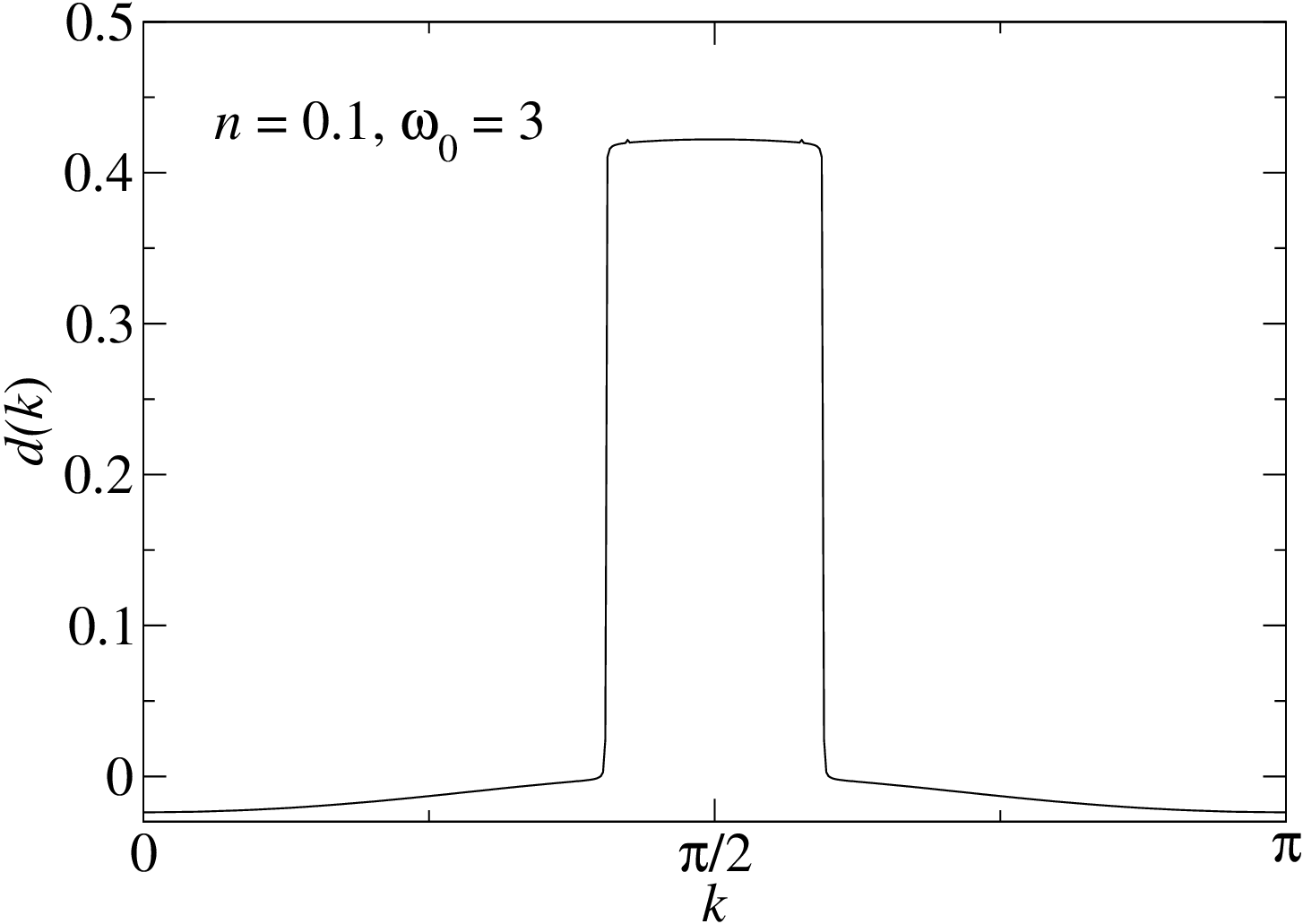}
\end{minipage} 
\caption{Bloch state occupation numbers $n(k)$ (left) and related correlation function $d(k)$ (right) for $n=0.9$ (upper panels) and 
$n=0.1$ (lower panels) at $\lambda=0$, $\omega_0=3$. 
Analytical results (lines) are compared
to numerical DMRG data (symbols).}
\label{fig_n_k_n09n01}
\end{figure}

In the case $n=0.1$ with $\omega_0/t_0 =3$ we find a self-consistent CDW state with order parameter $P=0.61$. The discrepancy in order of magnitude between this value and the much smaller one for $n=0.9$ is due to the consistent use of fermion density in the definition of $P$ (Eq.~(\ref{nP})), rather than changing to hole density for the $n=0.9$ case. For $n=0.1$ it has proved difficult to converge to a CDW solution in DMRG. This may indicate that the Green's function approximation is failing in this low density case. If so, the nature of the ground state is unclear. Nevertheless in Fig.~\ref{fig_spec_n01} we show results for the spectral function and in Fig.~\ref{fig_n_k_n09n01} we plot $n(k)$ and $d(k)$. The bottom of the narrow quasiparticle band near $E_F$ is at $k=\pi /2$ 
and only the majority subband is occupied. Hence $d(k)$ is strongly positive near $k=\pi /2$ and almost equal to $n(k)$. Over most of the zone the only contribution is from spectral weight further below the Fermi level; this vanishes at $k=\pi /2$ and makes a negative contribution to $d(k)$ owing to more weight residing on the minority sublattice. The discontinuities in $n(k)$ and $d(k)$ at the expected values $\pi/2 \pm \pi/10$ are clearly seen. 

\subsection{The CDW state at half-filling}\label{CDW}
We now consider a self-consistent CDW state for the half-filled band with $n=0.5$ and $\lambda=0$, $\omega_0 =3$. As before the Fermi energy 
$E_F$ (or chemical potential
$\mu$ at $T>0$) and order parameter $P$ are determined by Eqs.~\eqref{nP} and \eqref{sublattocc}, with the sublattice densities of states given by Eq.~\eqref{Nl}. For the above parameters we find $\mu=-0.211, P=0.765$. $E_{\rm F}$ lies in a gap as is appropriate for an insulator. 
\begin{figure}[b]
 \begin{minipage}{0.5\hsize}
  \centering
  \includegraphics[width=0.85\textwidth]{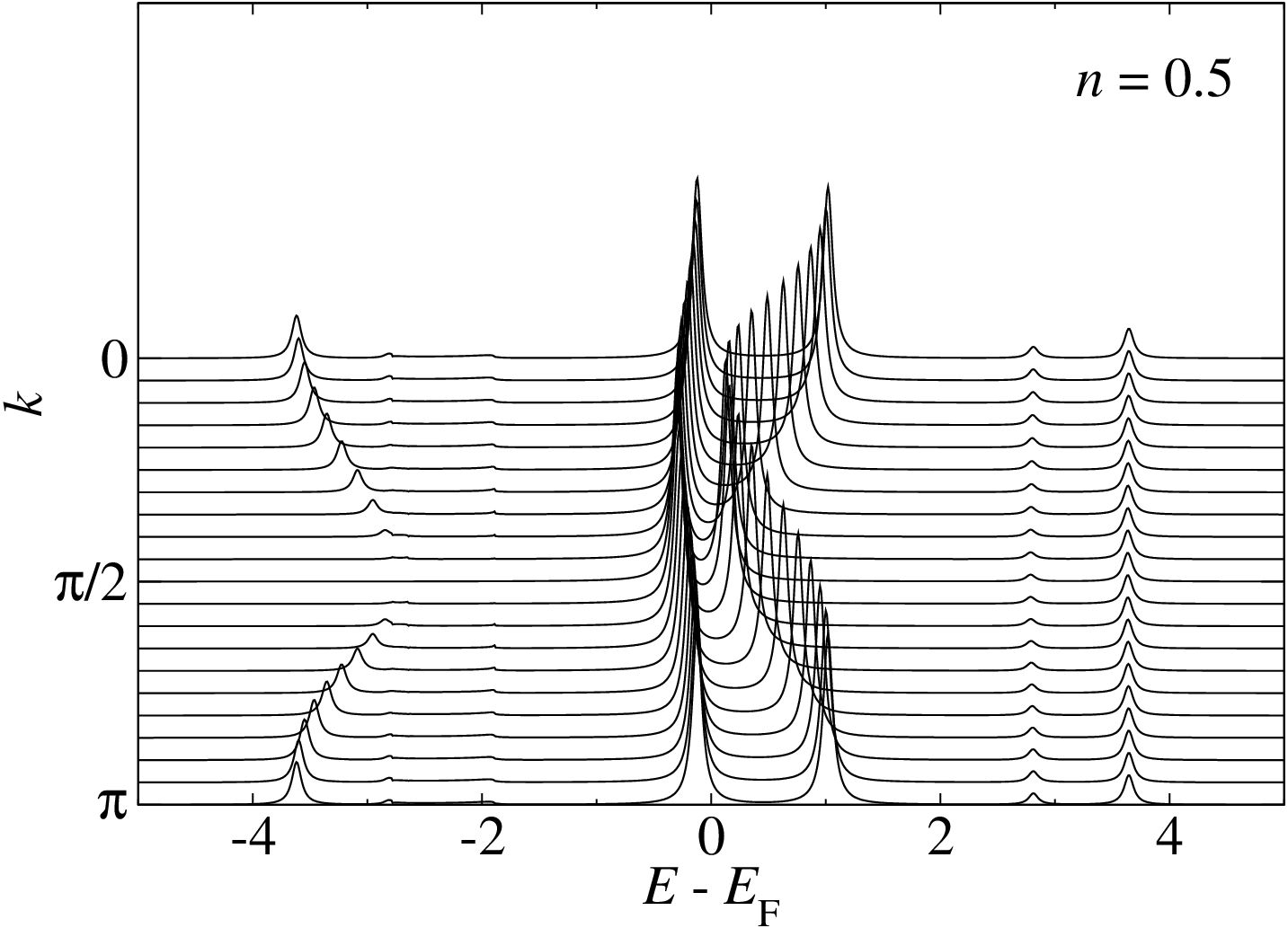}
 \end{minipage}
\begin{minipage}{0.5\hsize}
 \centering
 \includegraphics[width=0.85\textwidth]{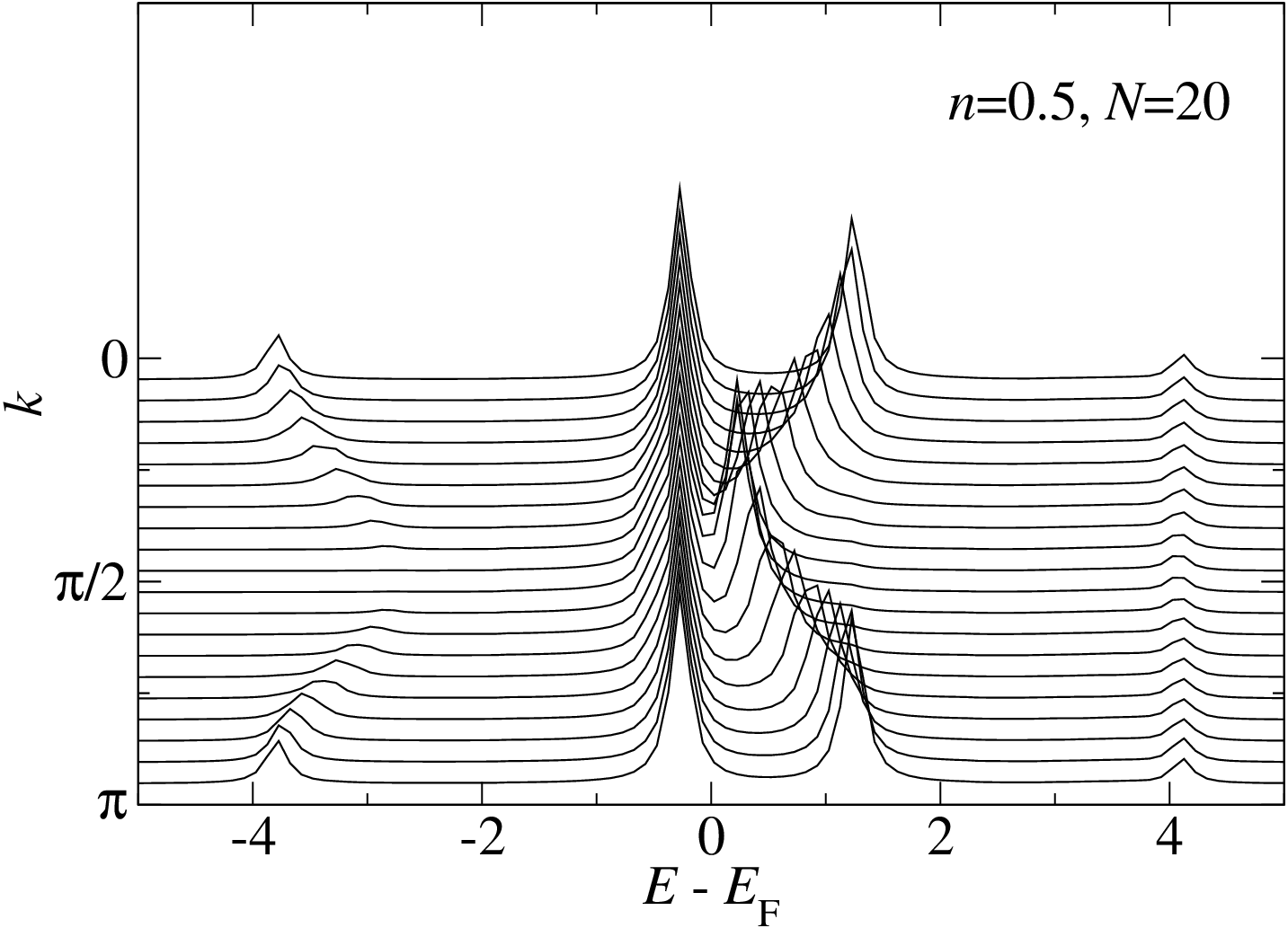}
\end{minipage} 
\caption{Single-particle spectral function $S(k,E)$ at $n=0.5$
 for $\lambda=0$, $\omega_0=3$ from Eq.~(\ref{totalspectralfunction}) (left panel)
 compared with the DDMRG result (right panel).
 }
\label{fig_spec_n05}
\end{figure}

\begin{figure}[b]
 \begin{minipage}{0.5\hsize}
  \centering
  \includegraphics[width=0.85\textwidth]{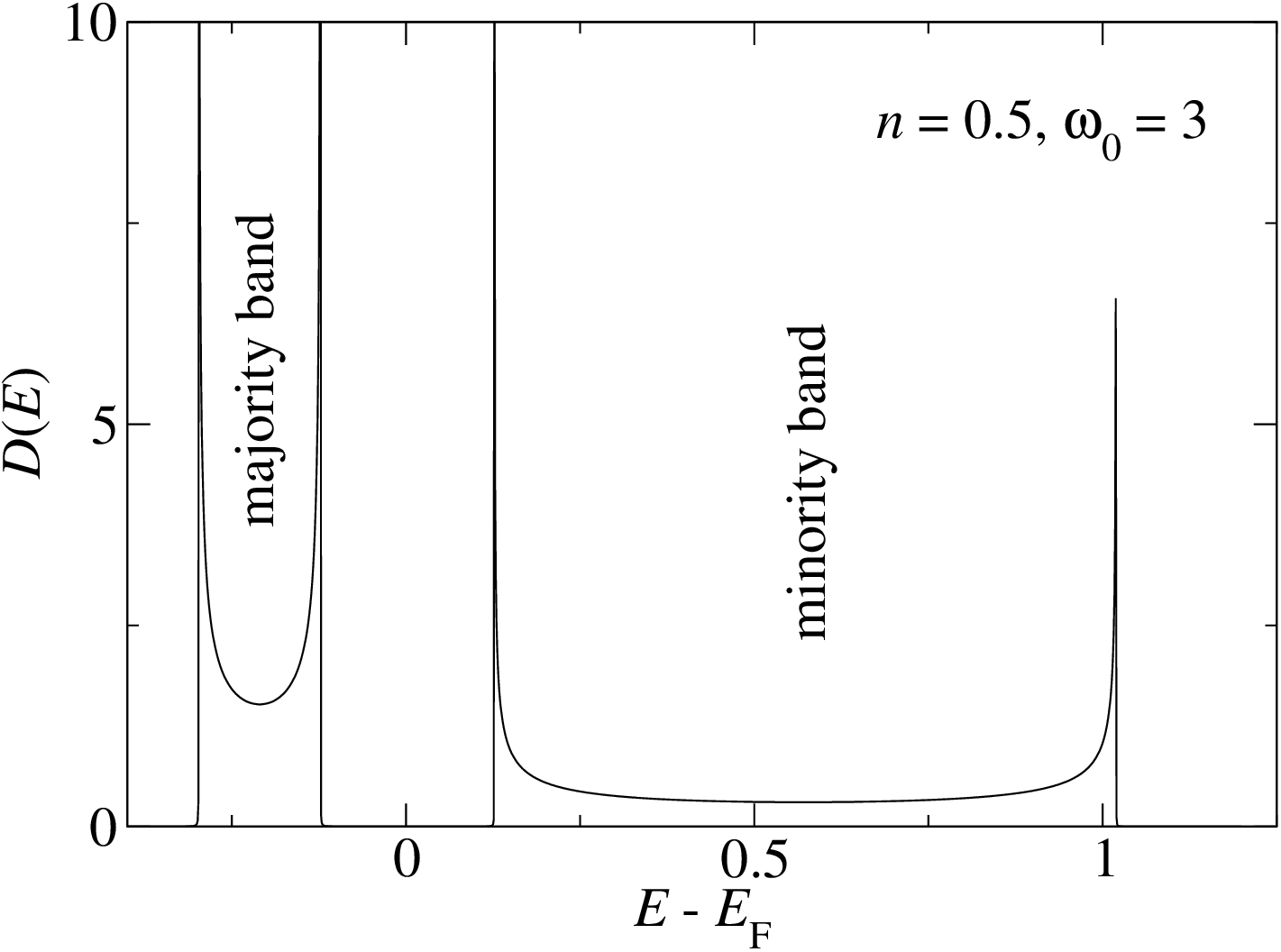}
 \end{minipage}
\begin{minipage}{0.5\hsize}
 \centering
 \includegraphics[width=0.55\textwidth]{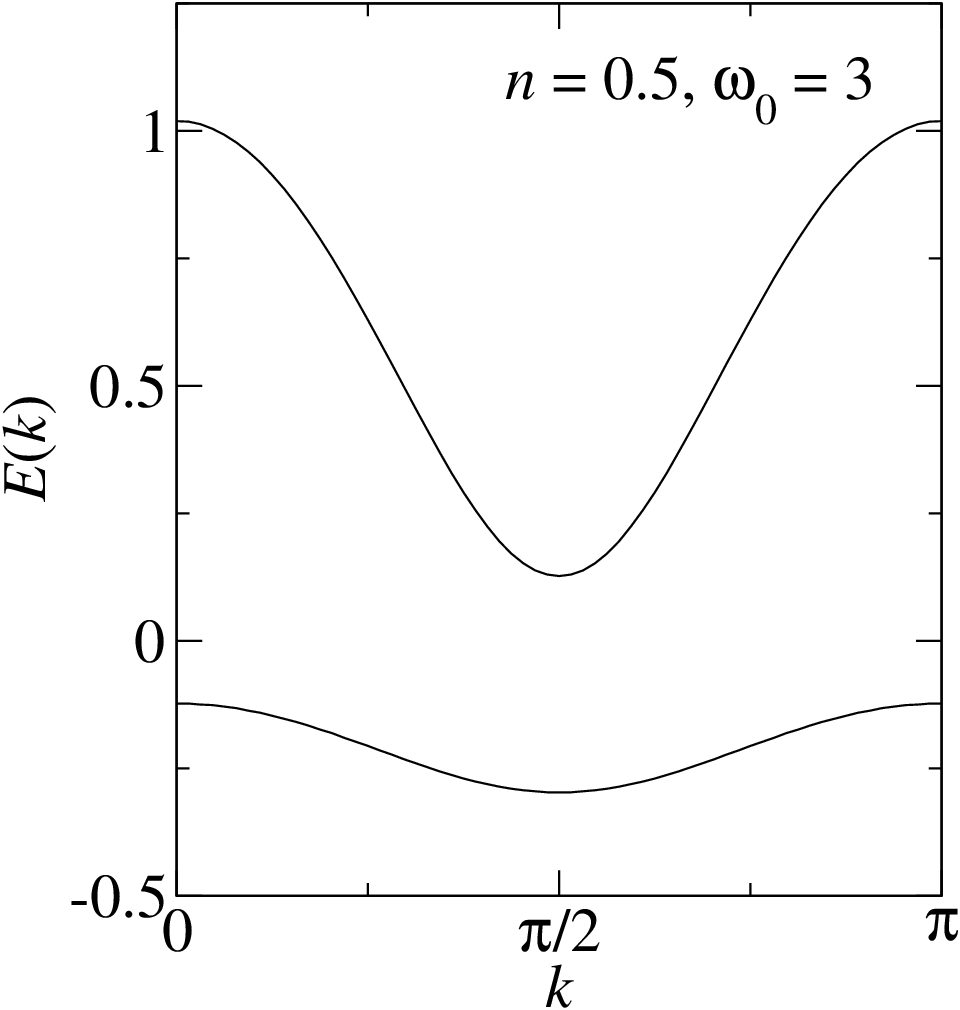}
\end{minipage} 
\caption{Density of states $D(E)$ near the Fermi level for $\lambda=0$
at half filling from Eq.~(\ref{Nl}) (left panel) and corresponding  
quasiparticle band dispersion $E(k)$ (right panel).}
\label{fig_Eknk_0.5}
\end{figure}

The spectral functions calculated from the Green's function method and the DDMRG are compared in Fig.~\ref{fig_spec_n05}. The agreement is generally good. The Fermi level lies in a gap between two quasiparticle bands, a broad upper unoccupied one and a narrower lower occupied one. The main discrepancy is that in the DDMRG case the lower quasiparticle band is extremely narrow whereas in the Green's function method it has a significant width. This shortcoming of the Green's function approximation is discussed thoroughly in Sec.~\ref{discussion}. Fig.~\ref{fig_Eknk_0.5} shows the density of states near the Fermi level, projected onto the majority and minority sublattices, calculated from Eq.~\eqref{sublatticedensityofstates}. It is remarkable that states in the occupied quasiparticle band are entirely confined to the majority sublattice, whereas those in the unoccupied quasiparticle band reside entirely on the minority sublattice. The dispersion curves $E(k)$ of the quasiparticle bands, obtained by plotting the loci of quasiparticle peaks in the spectral function, are shown in Fig.\ref{fig_Eknk_0.5}. The approximate Green's function therefore predicts an indirect gap, with the top of the occupied quasiparticle band at $k=0,\pi$ and the bottom of the unoccupied band at $k=\pi/2$.

\begin{figure}[t]
 \begin{minipage}{0.5\hsize}
  \centering
  \includegraphics[width=0.85\textwidth]{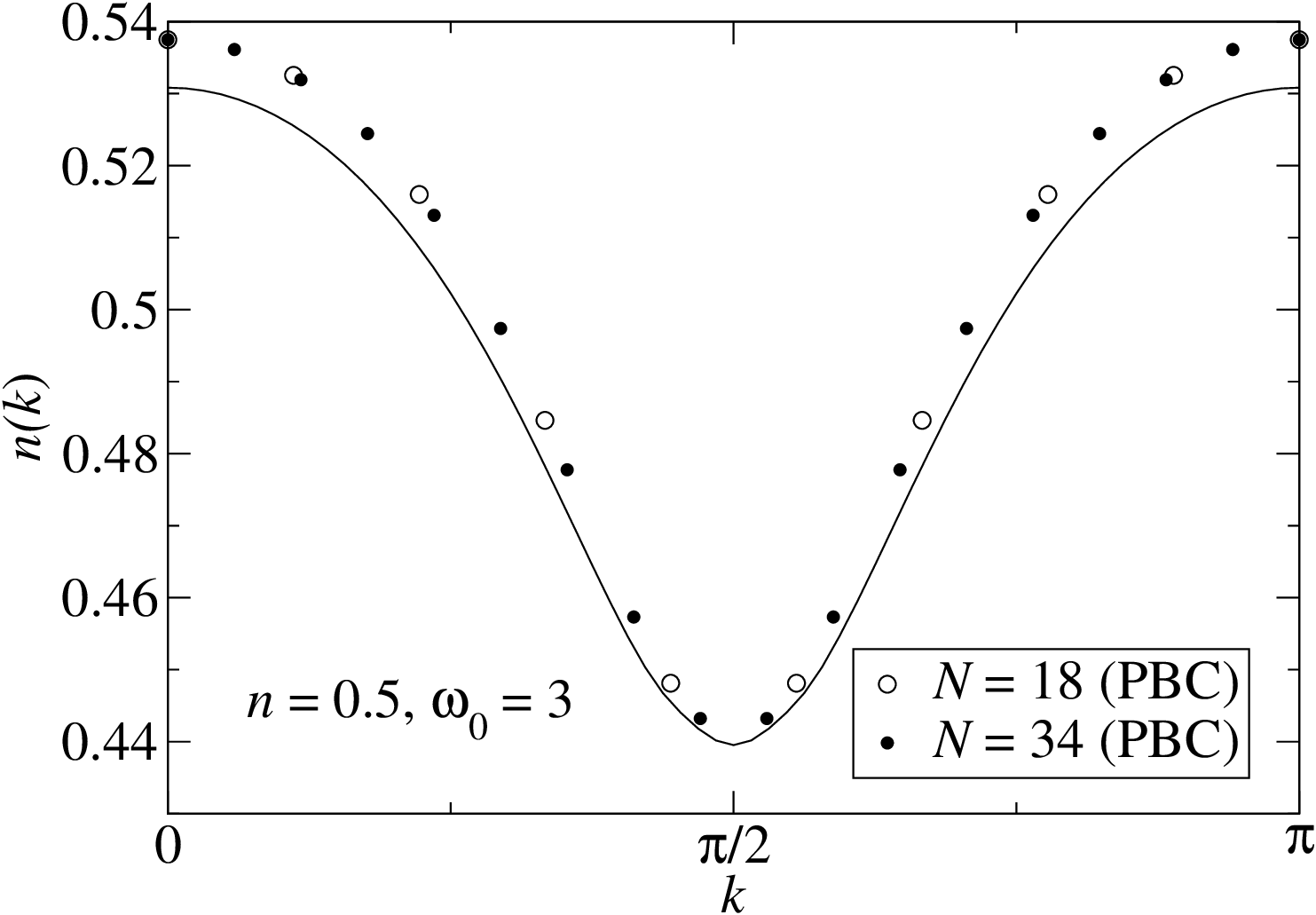}
 \end{minipage}
\begin{minipage}{0.5\hsize}
 \centering
 \includegraphics[width=0.85\textwidth]{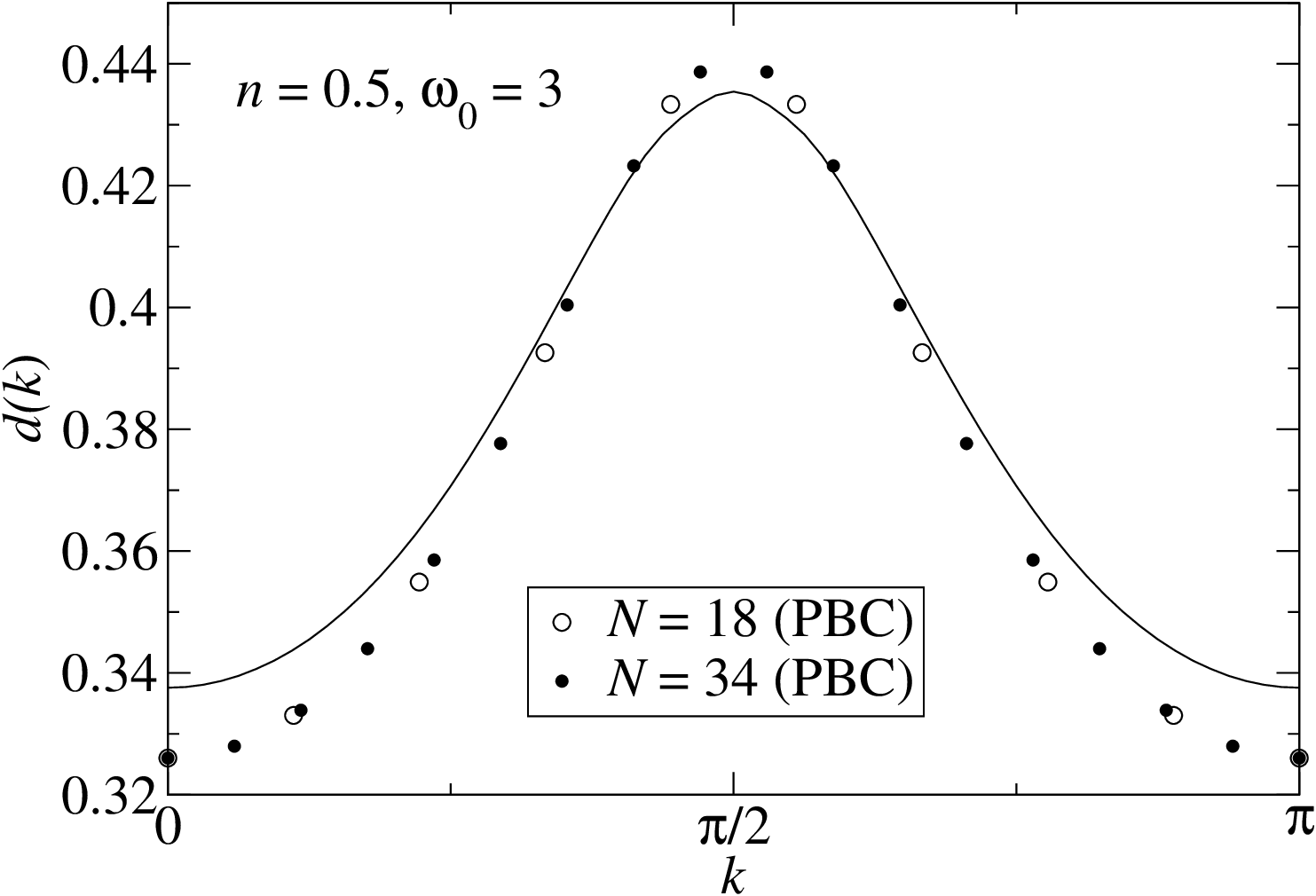}
\end{minipage} 
\caption{Bloch state occupation number $n(k)$ (left panel) and related
correlation function $d(k)$ (right panel) for the half-filled band case
with $\lambda=0$, $\omega_0=3$. 
 }
\label{fig_nkdk_n05}
\end{figure}
\begin{figure}[h]
 \begin{minipage}{0.5\hsize}
  \centering
  \includegraphics[width=0.85\textwidth]{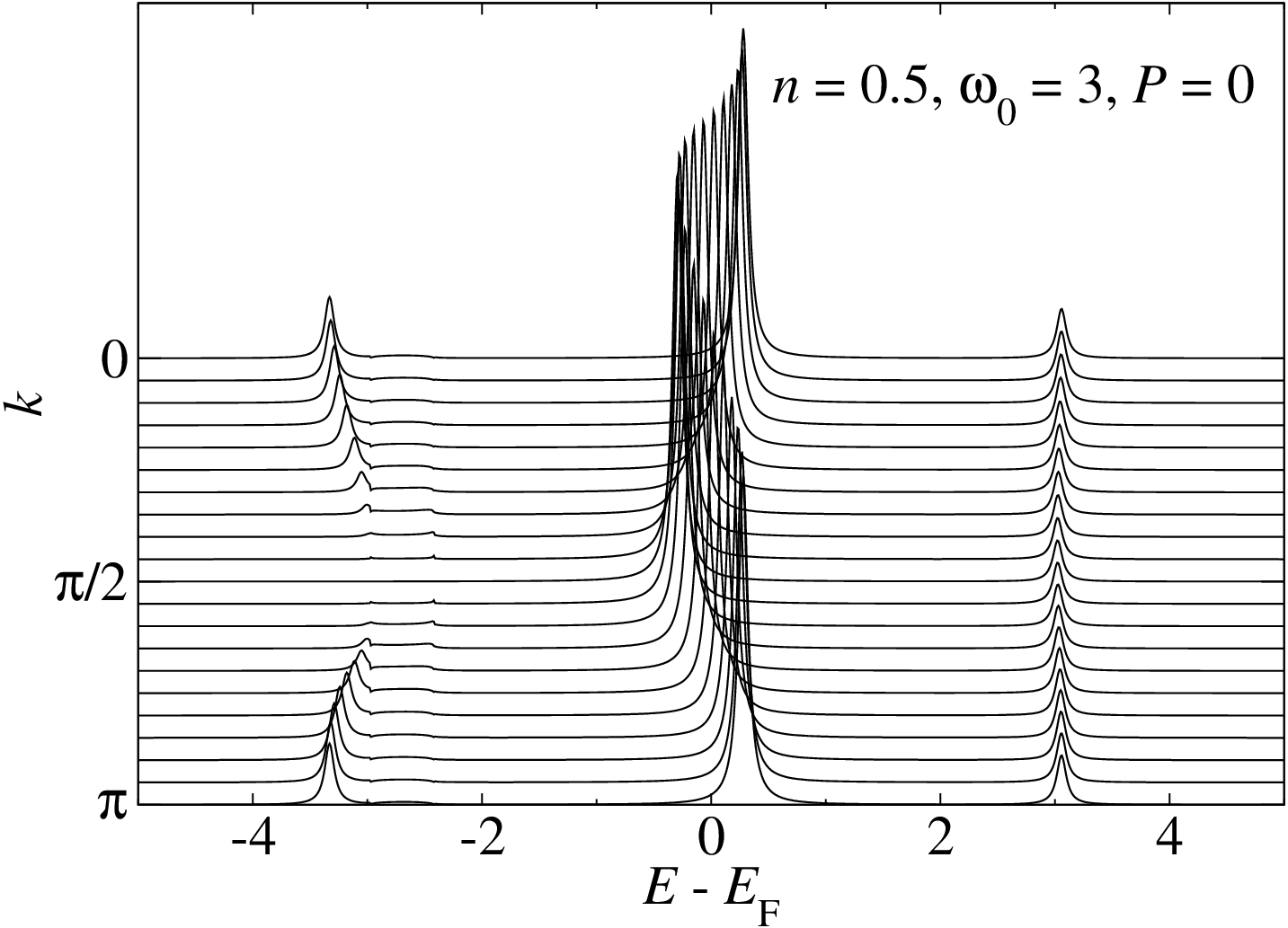}
 \end{minipage}
\begin{minipage}{0.5\hsize}
 \centering
 \includegraphics[width=0.55\textwidth]{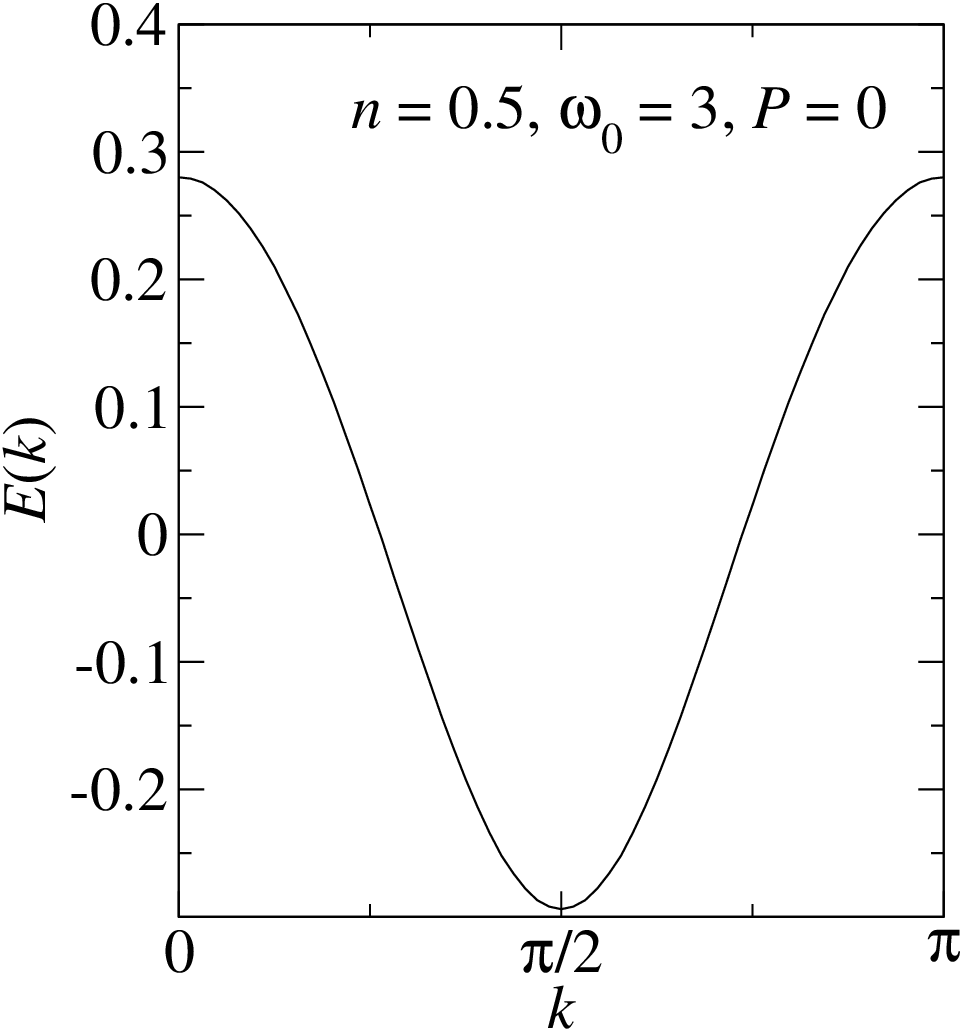}
\end{minipage} 
\caption{Single-particle spectral function (left panel) and quasiparticle band (right panel) in the high-temperature limit $P=0$ for the half-filled band case
with $\lambda=0$, $\omega_0=3$. 
 }
\label{fig_spec_P0}
\end{figure}
 In Fig.~\ref{fig_nkdk_n05} left and right panels  
we plot curves for $n(k)$ and $d(k)$, respectively,
 calculated from Eqs.~\eqref{nk} and~\eqref{dk}, together with points obtained by the DMRG method for finite systems of different sizes. The agreement is very striking, particularly since in DMRG these quantities are calculated directly from the ground state whereas in the Green's function method they are obtained as integrals over the spectral function. It should be noted that $n(\pi/2)=d(\pi/2)$ because for $k=\pi/2$ all the spectral weight of the occupied states resides in the quasiparticle peak which is entirely based on the majority sublattice. From Eqs.~\eqref{nk} and~\eqref{dk} it is clear that the weight in this peak is $2n(\pi/2)\simeq 0.88$.

Finally in Fig.~\ref{fig_spec_P0} we plot the spectral function (left panel)
and quasiparticle energy (right panel) in the high-temperature limit, where all short-range order has disappeared ($P=0$). The Fermi wave-vectors are close to $\pi/4$ and $3\pi/4$ with the correct Fermi surface 'volume'. This is the state from which the CDW evolves as the temperature is lowered. Clearly the situation is quite different from the usual one in which a CDW evolves because of nesting between Fermi wave-vectors at $k=\pm\pi/2$. The origin of this difference is 
that for $\lambda=0$ only next-nearest neighbour hopping, resulting in $\pi$-periodicity in $k$-space, occurs even in the disordered state.

\subsection{Discussion}\label{discussion}
The most notable difference between the Green's function and DDMRG results lies in the width of the occupied quasiparticle band for $n=0.5$. A very narrow band is also found in earlier calculations for $\omega_0 =2$ using exact diagonalisation where there is also no tendency for the top of the band to be at $k=0,\pi$ rather than $\pi/2$~\cite{WFAE08}. The band seems to be almost as narrow as one would have from the Trugman-like six-step process in a perfect CDW ($P=1$). This is quite surprising since for $\omega_0 =2$ the densities on the two sublattices are about 0.8 and 0.2~\cite{EHF09} which is far from a perfect CDW. 
Satellites below the main quasiparticle peak are suggestive of finite-size effects in the systems considered with 12 and 16 sites. 

To gain more insight into the Green's function approximation we may consider the limit of large $\omega_0$ where all the weight is concentrated in the two quasiparticle bands. From Eq.~\eqref{Gkmfinal} it follows that these bands are given by
\begin{equation}\label{Ek}
E(k)=\frac{\pm 2P+(1\pm P)\cos(2k)}{\omega_0}\,,
\end{equation} 
where the upper signs $(+)$ correspond to the upper band and the lower signs $(-)$ to the lower band. The widths $2(1\pm P)/\omega_0$ of these bands can be understood by inspection of the following processes, where a bullet represents a vacant site, a star represents a boson and a circle represents a fermion:
\begin{equation*}\label{fermionhop}
\vert\,\bullet\;\bigcirc\;\bigcirc\rangle\rightarrow\vert\bigcirc\;\bigstar\;\bigcirc\rangle\rightarrow
\vert\bigcirc\,\bigcirc\;\bullet\,\rangle
\end{equation*}
\begin{equation*}\label{holehop}
\,\vert\bigcirc\,\bigcirc\;\bullet\,\rangle\rightarrow\vert\bigcirc\;\bigstar\;\bigcirc\rangle\rightarrow
\vert\,\bullet\;\bigcirc\;\bigcirc\rangle
\end{equation*}
 In the upper line the central site is on the majority sublattice. The diagram shows how a fermion added to the upper band, on the minority sublattice, can hop by a two-step process across an occupied majority site. The probability of the majority site being occupied is $(1+P)/2 $ so an estimate of the width of the upper band is $2(1+P)/\omega_0$, as in Eq.~\eqref{Ek}. In the lower line the central site is on the minority sublattice. Clearly a hole created on the majority sublattice can hop by a two-step process across an occupied minority site. The probability of the minority site being occupied is $(1-P)/2$ so an estimate of the width of the lower quasiparticle band is $2(1-P)/\omega_0$ as in Eq.~\eqref{Ek}. The probability argument used here is equivalent to the Hartree-Fock-like approximation made in deriving the Green's function. 
This approximation is clearly failing in the present situation, since the width of the lower quasiparticle band is much larger than that given by the DDMRG. To expose the cause of this failure we used DMRG to calculate the three-site correlation functions $\langle n_{m-1}n_{m} n_{m+1} \rangle$ for a 10-site system with periodic boundary conditions and $\lambda=0$, $\omega_0 /t_0 =3$ as usual. The correlation function takes two values $6.46062\times 10^{-4}$ and $3.90826\times 10^{-3}$, depending on whether $m$ is an even or odd site corresponding to the majority and minority sublattice, respectively. The simple probability argument would give respective values $P(1-P)^2 =0.04225$, $(1-P)P^2 =0.1375$ when $P=0.765$. If a site on the minority sublattice is occupied it is clear that the probability of both neighbouring sites being occupied is only about $0.039$. Thus one of these majority sublattice sites is very likely to be occupied by a hole. Hence there is a strong tendency for the minority sublattice fermions and the majority sublattice holes, equal in number, to form bound pairs. We conclude that for $n=0.5$ important correlations exist in the ground state which are not included in the Hartree-Fock-like approximation of the Green's function decoupling. 

\section{Conclusions and outlook}\label{conclusions}
Doped Mott insulators remain at the forefront of physics research largely
because of their relevance to high temperature superconductivity. In the
cuprate superconductors holes in the copper oxide planes move in a background
of antiferromagnetic order. In other systems of interest the background is
one of alternating orbital order. The progress of a hole through such a background is hindered by the string effect. This effect has been known for a 
long time~\cite{BR70}
but the string picture is central to much recent work 
(e.g.~\cite{WSE08,WOH09,Be09,WO10}).
The Edwards fermion-boson model considered in this paper 
was introduced to describe
this effect in the simplest possible way~\cite{Ed06}. The spinless fermions
correspond physically to the holes in the Mott insulator. The ordered background
does not appear explicitly in the model; the essence of the string effect
actually relies only on the existence of substantial short-range order without
the necessity of true long-range order. Clearly the physical interest  
lies mainly in 2D~\cite{BF10}, but 
so far most calculations for the Edwards model have been made in 
1D~\cite{AEF07,WFAE08,EHF09,EF09b,AEF10,SBF10}.
In this paper we describe an analytical approximation to the one-fermion
Green's function which is valid in 1D, 2D and 3D. Its main limitations are
that the boson energy should be fairly large  ($\omega_0/t_0>2$) and that
string relaxation is neglected ($\lambda=0$). The principal objective of
this paper is to test the accuracy of the Green's function method, within
its expected domain of validity, by comparing with numerical results obtained
in 1D by the DMRG and DDRMG methods.

This comparison has been made in detail for $\omega_0 =3$ and for various
fermion densities. For the half-filled band case ($n=0.5$) excellent agreement
is obtained for ground state properties. These include the CDW order parameter,
the Bloch state occupation number $n(k)$ and a related quantity $d(k)$ associated
with the CDW. There is also generally good agreement for the one-fermion spectral function although the Green's function method predicts much too wide an occupied quasiparticle band. In Sec.~\ref{discussion} the reason for this discrepancy is traced to missing correlations in the Green's function approximation. 
The Green's function method predicts that the CDW
state evolves from a high-temperature disordered state with Fermi wave-vectors
close to $\pi/4$ and $3\pi/4$, which is quite different from the usual case
where a CDW evolves because of nesting between Fermi wave-vectors at $k=\pm
\pi/2$. The origin of the difference is that only next-nearest neighbour
hopping, resulting in $\pi$-periodicity in $k$-space, 
occurs even in the disordered state.

There is excellent agreement between the two methods for the dilute 
hole ($n=0.9$) case with the usual parameters $\lambda=0$, 
$\omega_0 /t_0 =3$. Somewhat surprisingly, there is also a 
two-sublattice CDW state which in this case is metallic. 
In the Appendix  it is shown how the appearance of 
two-sublattice CDW states is related to a symmetry property of the model 
with $\lambda=0$.
The comparison between the two methods proved to be more difficult in 
the dilute fermion ($n=0.1$) case. The Green's function method again 
predicts a metallic CDW state but the DMRG fails to confirm this. 
The true nature of the ground state in this case remains unclear.

It may be concluded that the rather simple Green's function approximation
derived here is sufficiently successful in 1D, which is probably the least
favourable case, to envisage future applications to the 2D 
$\mathrm{t-J_z}$ model and, with a slight extension of the model, to the 
$\mathrm{t_{2g}}$ model of alternating orbital order.  The situation of physical interest will be low to moderate fermion density, the fermions corresponding to holes in the relevant Mott insulator.

\section*{Acknowledgements} This work was supported by SFB 652 of the 
Deutsche Forschungsgemeinschaft. 
\section*{Appendix}
Through the particular form of the fermion-boson hopping term at
$\lambda=0$ (cf. Eq.~(\ref{EdwardsHamBar})  
the transfer of a fermion beween neighbouring lattice sites 
coincides with a
change of the number of bosons by one. As a consequence we may think of
the bosons as tracking the motion of the fermions. This picture becomes
exact through the idenfication
of a conserved quantity.
Let us define operators $N_f^A$ and $N_f^B$ which count the 
number of fermions on the A- or B-sites of a bipartite lattice, 
and similar operators $N_b^A$ and $N_b^B$ for bosons.
Then $N_{fb} = N_f^A - N_f^B + 2(N_b^A - N_b^B)$  commutes with the 
Hamiltonian $\tilde{H}_{Ed}$, when $\lambda=0$, and is therefore a conserved quantity.
The eigenvalues of $N_{fb}$ can be used to classify the eigenstates 
of the Hamiltonian. We note that $N_{fb}$ is not conserved for 
$\lambda \neq 0$.

The existence of $N_{fb}$ has two major consequences.
First, fermion operators such as  $f_{nA}^{\dagger}f_{mB}^{}$ change  
$N_{fb}$ by $2$, so that in any eigenstate of the Hamiltonian the expectation value $\langle f_{nA}^{\dagger}f_{mB}^{} \rangle=0$ for arbitrary sites 
$nA$ and $mB$ on the respective sublattices. 
Similarly  $\langle b_n \rangle=0$.
Hence quantities such as the spectral function $S({\bf k},E)$ or
the momentum distribution $n({\bf k})$ have the periodicity in ${\bf k}$ 
of the reciprocal lattice of the real-space A or B sublattice, 
e.g. $\pi$-periodicity in the 1D case.
Second, it implies that those eigenvalues of the Hamiltonian 
corresponding to $N_{fb} \neq 0$  are degenerate.
This follows because the translation operator $T$ commutes with $H$, but 
changes the sign of $N_{fb}$. The energy eigenvalue of an eigenstate
$|\psi\rangle$ with $\langle \psi |N_{fb}|\psi \rangle \neq 0$ 
must therefore be (at least) two-fold
degenerate, since the state $T|\psi\rangle$ belongs to the same
energy but differs from $|\psi\rangle$ due to the change of $N_{fb}$.
 This degeneracy corresponds to a breaking of translational
symmetry, as for a two-sublattice CDW state.
Clearly if the ground state at $\lambda = 0$ is not such a CDW state it
must have $N_{fb} = 0$.
\section*{References}

\begin{thebibliography}{10}
\expandafter\ifx\csname url\endcsname\relax
  \def\url#1{{\tt #1}}\fi
\expandafter\ifx\csname urlprefix\endcsname\relax\def\urlprefix{URL }\fi
\providecommand{\eprint}[2][]{\url{#2}}

\bibitem{Hu63}
Hubbard J 1963 {\em Proc. Roy. Soc. London, Ser. A\/} {\bf 276} 238

\bibitem{DWOAH08}
Daghofer M, Wohlfeld K, Ole\'{s} A~M, Arrigoni E and Horsch P 2008 {\em Phys.
  Rev. Lett.\/} {\bf 100} 066403

\bibitem{WDOH08}
Wohlfeld K, Daghofer M, Ole\'{s} A~M and Horsch P 2008 {\em Phys. Rev. B\/}
  {\bf 78} 214423

\bibitem{MOKT05}
Matsuno J, Okimoto K, Kawasaki M and Tokura Y 2005 {\em Phys. Rev. Lett.\/}
  {\bf 95} 176404

\bibitem{HIYW83}
Hidaka M, Inoue K, Yamada I and Walker P~J 1983 {\em Physica B \& C\/} {\bf
  121} 343

\bibitem{MDTTBPSB06}
McLain S~E, Dolglos M~R, Tennant D~A, Turner J~F~C, Barnes T, Proffen T, Sales
  B~C and Bewley R~I 2006 {\em Nature Mater.\/} {\bf 5} 561

\bibitem{CSO77}
Chao K~A, Spalek J and Ole\'{s} A~M 1977 {\em J. Phys. C\/} {\bf 10} L271

\bibitem{BR70}
Brinkman W~F and Rice T~M 1970 {\em Phys. Rev. B\/} {\bf 2} 4302

\bibitem{KLR89}
Kane C~L, Lee P~A and Read N 1989 {\em Phys. Rev. B\/} {\bf 39} 6880

\bibitem{Tr88}
Trugman S~A 1988 {\em Phys. Rev. B\/} {\bf 37} 1597

\bibitem{MH91a}
Martinez G and Horsch P 1991 {\em Phys. Rev. B\/} {\bf 44} 317

\bibitem{Ed06}
Edwards D~M 2006 {\em Physica B\/} {\bf 378-380} 133

\bibitem{Sti73}
Stinchcombe R~B 1973 {\em J. Phys. C\/} {\bf 6} 2459

\bibitem{AEF07}
Alvermann A, Edwards D~M and Fehske H 2007 {\em Phys. Rev. Lett.\/} {\bf 98}
  056602

\bibitem{WFAE08}
Wellein G, Fehske H, Alvermann A and Edwards D~M 2008 {\em Phys. Rev. Lett.\/}
  {\bf 101} 136402

\bibitem{EHF09}
Ejima S, Hager G and Fehske H 2009 {\em Phys. Rev. Lett.\/} {\bf 102} 106404

\bibitem{EF09b}
Ejima S and Fehske H 2009 {\em Phys. Rev. B\/} {\bf 80} 155101

\bibitem{Wh92}
White S~R 1992 {\em Phys. Rev. Lett.\/} {\bf 69} 2863

\bibitem{Wh93}
White S~R 1993 {\em Phys. Rev. B\/} {\bf 48} 10345

\bibitem{BHS02}
Becker K~W, H\"ubsch A and Sommer T 2002 {\em Phys. Rev. B\/} {\bf 66} 235115

\bibitem{SBF10}
Sykora S, Becker K~W and Fehske H 2010 {\em Phys. Rev. B\/} {\bf 81} 195127

\bibitem{AEF10}
Alvermann A, Edwards D~M and Fehske H 2010 {\em J. Phys. Conf. Ser.\/} {\bf
  220} 012023

\bibitem{Zu60}
Zubarev D~N 1960 {\em Usp. Fiz. Nauk\/} {\bf 71} 71

\bibitem{Je02b}
Jeckelmann E 2002 {\em Phys. Rev. B\/} {\bf 66} 045114

\bibitem{WSE08}
Wr\'{o}bel P, Suleja W and Eder R 2008 {\em Phys. Rev. B\/} {\bf 78} 064501

\bibitem{WOH09}
Wohlfeld K, Ole\'{s} A~M and Horsch P 2009 {\em Phys. Rev. B\/} {\bf 79} 224433

\bibitem{Be09}
Berciu M 2009 {\em Physics\/} {\bf 2} 55

\bibitem{WO10}
Wr\'{o}bel P and Ole\'{s} A~M 2010 {\em Phys. Rev. Lett.\/} {\bf 104} 206401

\bibitem{BF10} Berciu M and Fehske H 2010  {\em Phys. Rev. B\/} {\bf 82} 085116

\end{thebibliography}
\providecommand{\newblock}{}

\end{document}